\documentclass[10pt,aps,prx,twocolumn,showpacs,notitlepage,floatfix,superscriptaddress,longbibliography]{revtex4-2}
\usepackage{hyperref}
\hypersetup{
colorlinks=false,
colorlinks=true,
citecolor=blue,
linkcolor=blue,
urlcolor=blue}
\usepackage{graphicx}
\usepackage{amsmath}
\usepackage{physics}
\usepackage{amsfonts}
\usepackage{amssymb}
\usepackage{mathtools}
\usepackage{bm}
\usepackage{siunitx}
\usepackage{xfrac}
\usepackage[normalem]{ulem}
\usepackage{orcidlink}  
\usepackage{braket}
\usepackage{cases}
\usepackage[dvipsnames, svgnames]{xcolor}

\newcommand{\mdas}[1]{\textcolor{black}{#1}}

\DeclareUnicodeCharacter{0229}{\c{e}}

\begin{document}

\title{Dynamic hysteresis and transitions controlled by asymmetry in potential: energetic and entropic aspects}

\author{Samudro Ghosh}
\email[]{Contact author: d22065@students.iitmandi.ac.in}

\author{Moupriya Das\orcidlink{0000-0003-1851-2162}}
\email[]{Contact author: moupriya@iitmandi.ac.in}
\affiliation{Indian Institute of Technology Mandi, Kamand, Himachal Pradesh 175075, India
}

\begin{abstract}

We aim to investigate the precise role of the asymmetry in the underlying potential on the process of dynamic hysteresis. So far, the theoretical modeling of this phenomenon of fundamental importance has been done considering symmetric bistable systems. The influence of asymmetry in the bistable potential has been explored in related contexts, such as first passage time estimates, escape dynamics, stochastic resonance, etc., highlighting the importance of the aspect of asymmetry in the systems. However, the perspective of developing a thorough theoretical understanding in this context remained overlooked in the case of dynamic hysteresis. Here, we scrutinize how the asymmetry in the potential influences this process, considering the physical model of a Brownian particle in a double-well potential subject to a periodic forcing. We analyze in detail the effect of two distinct types of asymmetry in the intrinsic potential, one corresponding to the different depths and the other subject to the separate widths of the two wells of the double-well potential representing the bistable system. Our study reveals that the hysteretic effect predominantly decreases with increasing asymmetry in both types of the system, reflected in the reduced hysteresis loop area. This observation suggests that the introduction of the appropriate asymmetry in the potential can quantitatively regulate the outputs of the dynamic hysteresis. This modulation can be beneficial for controlling devices' outputs in which dynamic hysteresis plays a role. Moreover, importantly, the implemented asymmetries in the potential can induce symmetry breaking in the response of the systems. This gives rise to asymmetric dynamic hysteresis loops, signifying the dynamic transition to an asymmetric phase, in moderate conditions, which is absent in the symmetric systems in the same scenarios.

\end{abstract}

\maketitle

\section{Introduction}

\mdas{The behavior of the natural and designed systems depends on the factors 
governing their underlying dynamics, the regulation by external controls, if any exist, 
and the effect of the surroundings. 
Hysteresis is an important phenomenon where this behavior or response of the system depends 
on its past history, determined by the controlling components. 
Dynamic hysteresis is a class of hysteresis that emerges in the presence 
of an external periodic field~\cite{chakrabarti1999}. 
The system's feedback lags behind the external periodic control due to relaxational delay, 
giving rise to response function-field hysteresis loops. 
The effect of this hysteresis tends to vanish in the quasi-static limit, 
i.e., in the zero frequency extent of the external force. 
This is because, in this limiting condition, the response of the system can follow 
the extrinsic perturbation almost exactly, without any delay, because of the 
infinitely slow-changing nature of this periodic modulation. 
This reflects through the vanishing area of the hysteresis loops. 
This is distinct from static hysteresis, where the impact of hysteresis 
in the system's response is retained even in the quasi-static limit. 
For example, in pure magnetic materials, the magnetization exhibits 
dynamic hysteresis in response to the external magnetic field. 
The magnetization-field hysteresis loops tend to disappear for extremely slow-varying 
magnetic fields~\cite{landau1935, feynman1965, kittel1966}.
Whereas the magnetic systems with random pinning defects 
bear the effect of static hysteresis~\cite{strogatz1990}.
This is because here the memory effect in the system's behavior 
is developed as a consequence of the system's internal structure rather 
than by the influence of the external drive.
}\\ 

\mdas{Here, we mention that the functions of some engineered materials involve
static hysteresis. However, the majority of the designed engineering problems related to 
recording, memory device materials, etc., are based upon the working principle of
dynamic hysteresis~\cite{torre1966}. 
The manifestation of dynamic hysteresis is not just limited to magnetic materials.  
It has a wide presence in the operations of certain electrical machines, such as 
power transformers, electrochemical systems like batteries and supercapacitors, 
thermal sensors, actuators, shape-memory alloys, 
mechanical shock and seismic dampers, to mention a few. 
Also, it is observed in nature, for example, in some particular climate dynamics 
like forest-desertification, thermohaline flow in the ocean, 
and in many biological systems, which are subject to external periodic stimuli. 
Therefore, the study of dynamic hysteresis~\cite{suen1999, graham2005, fologea2011, das2012, das2012b, das2012c, banerjee2015, bonilla2015, romensky2015, kostanyan2020, anand2020, kundu2023} 
is considered to have immense significance 
in the fundamental understanding of the hysteresis mechanism as well as 
in the construction of the technological devices. 
} 


\mdas{The investigation on the vast range of systems where the dynamic hysteresis plays a role 
suggests that this dynamical phenomenon can have both advantageous and disadvantageous effects 
on the outcomes, depending on the systems under study. For example, in magnetic data storage 
devices~\cite{katzgraber2006, asakawa2009, moree2023}, the dynamic hysteresis plays a beneficial 
role because it allows the system to store information reliably by producing a `memory effect' 
in response to the applied magnetic field. Also, it provides immunity towards noise by 
restricting unwanted switching between the states. A similar kind of mechanism, through 
the prevention of the rapid, unwanted switching between accessible states, retention of memory 
and implementation of robustness in response to the presence of noise, 
acts beneficially in the functions of many devices, such as thermostats~\cite{zou1999, devergne2024}, latching relays~\cite{uchino1998}, Schmitt triggers~\cite{khovanov2007, khovanov2008, zhang2019}, etc.   
The efficient energy dissipation and the support of controllable damping through 
the process of dynamic hysteresis helps in the proper operations of 
magnetorheological~\cite{dominguez2006} and seismic dampers~\cite{putignano2016, rinaldin2017}, 
shape-memory alloys~\cite{qiao2011, xu2023}, etc. Moreover, this dynamical process 
has been found to have positive effects in a number of biochemical systems~\cite{pomerening2003, byrd2019, hartich2021, sasai2025}, 
including cell cycles and gene regulatory networks. In these cases, the advantages in the 
activities result from the fluctuations-resistant, irreversible transitions with memory 
achieved through dynamic hysteresis, which help in the proper decision-making 
in the biochemical networks. 
On the other hand, dynamic hysteresis bears disadvantageous effects in flip-flop circuits~\cite{djibaoui2025}
and sensor devices~\cite{hu2013, urs2020}
by introducing history and rate-dependent response. 
The delay in the generation of the feedback caused by dynamic hysteresis degrades the 
functions of these systems where high temporal precision is desirable. 
In devices like transformers~\cite{pong2009, rajnak2014, huang2024},
motors~\cite{gao2023},
etc., the non-beneficial role of dynamic hysteresis manifests primarily as the 
energy loss as the dissipated heat. In the actuator operations in robotics~\cite{bihler2008},
the lag in the response due to hysteresis produces lower accuracy in the output. 
Also, it causes the loss of the input energy in the form of heat. 
In many biological systems, the appropriate functions 
are based on real-time sensing without any delay, fast and precise switching, and unique or 
synchronized input-output correspondence. The cases include sensors~\cite{aydemir2010, mcgrath2014}
and neurons~\cite{feudel2018},
ion channels~\cite{das2012c, conroy2012, ding2020},
heart cells~\cite{huang2010, landaw2017},
etc. Here, the presence of dynamic hysteresis plays unfavorable roles. 
Dynamic hysteresis has also been identified as disadvantageous in some specific climate systems where it leads to the delayed recovery of the Earth system, for example, in the forest and desertification dynamics~\cite{feudel2018, consolo2022, pal2022, kaszas2016, kronke2020}
.}


\mdas{The above discussions suggest that the process of the dynamic hysteresis phenomenon can 
have a positive or a negative impact in a specific situation. Therefore, it would be a 
fundamental goal to understand the control over the extent of hysteresis. 
With this interpretation, the performance of the systems, the actions of which involve this 
dynamic process, can be improved by appropriate modulation of the degree of hysteresis. 
For this purpose, the development of a theoretical framework, modeling the mechanism 
of dynamic hysteresis becomes important. The implementation of the model can suggest 
the appropriate tuning of the degree of hysteresis by determining the controlling factors 
and their quantitative significance on the hysteretic effect. This theoretical understanding 
can be translated into real systems for practical applications.}


\mdas{Dynamic hysteresis is generally observed in many 
pure extensive cooperative systems due to the action of the 
external periodic field. However, it is most common and 
vastly studied in the domain of magnetic systems. The 
phenomenon can be understood as the interplay between two 
competing time scales related to the system and the 
dynamics; the relaxation time of the system and 
the time period of oscillation of the external field. 
In an early study, the dynamic process was interpreted in a 
driven bistable model system subject to rate competition~\cite{agarwal1981}. 
The idea was later followed in a multi-component magnetic model~\cite{rao1989, rao1990}.
In the majority of circumstances, the theoretical modeling for the cooperative 
magnetic system exhibiting dynamic hysteresis involves a 
mean-field approach\cite{chakrabarti1999}.
However, the classic mean-field approximation 
does not include the effect of the environmental 
fluctuations and therefore, is unable to capture the 
actual picture of the process\cite{chakrabarti1999}.
To take into account the consequences of the 
thermal fluctuations, several numerical simulation studies 
were done within the Monte Carlo framework\cite{chakrabarti1999}.
Later, the bistable Langevin model with an underlying 
double-well potential and subject to external periodic 
control was introduced to analyze the mechanism of dynamic 
hysteresis~\cite{mahato1994, paniconi1997}.
We point out that this setup incorporates all 
essential components for dynamic hysteresis to occur in the 
system: bistability, external periodic drive, 
and thermal noise. Therefore, it can appropriately justify 
all the features related to the process. Also, this model 
is not just limited to explaining dynamic hysteresis in 
the magnetic systems. It can be considered as a general 
framework to study dynamic hysteresis in the presence of 
elemental bistability and fluctuations in the surroundings. 
The areas range from analog circuits~\cite{gammaitoni1998},
ferroelectric~\cite{klotins2005} 
and piezoelectric materials~\cite{smith2003},
gene networks~\cite{hasty2000},
to climate systems~\cite{gammaitoni1998, das2020}.  
We refer to the fact that the same model can be potentially 
applied to study other important related phenomena like stochastic 
resonance~\cite{gammaitoni1998},
resonant activation~\cite{doering1992}, 
driven escape problems~\cite{wang2017, cilenti2022} 
and periodically driven stochastic systems, in general~\cite{jung1993}.}


\mdas{It becomes apparent that the described Langevin 
dynamics model is an effective theoretical setup to 
elucidate many significant dynamical processes. 
It was employed successfully to specify the quantitative  
control over the extent of dynamic hysteresis by the  
relevant factors linked to the system and dynamics . The nontrivial dependence of the 
hysteretic effect on the fluctuations or noise strength and 
the driving frequency was revealed~\cite{mahato1994}.
Here, we point out that the influences 
of the parameters related to the external driving and the 
thermal fluctuations in the process of dynamic hysteresis 
have been analyzed thoroughly. However, the precise effect 
on the mechanism through another crucial component, 
the underlying potential, remained unexplored so far. 
In the studies related to this phenomenon, the potential is 
conventionally considered to have the symmetric bistable 
form to reflect the elementary bistability~\cite{vemuri1989, mahato1994}. 
In the present study, we aim to explore this domain 
comprehensively, which appears not to 
have been focused on yet. More specifically, we address the 
fundamental role of the asymmetry in the structure of the 
underlying potential on dynamic hysteresis. 
We scrutinize in detail how the degree of asymmetry in the 
potential regulates the extent of the outcome of this 
dynamic process.}


\mdas{We emphasize that while the consequences of asymmetry 
in potential have been analyzed in the case of the related 
processes, such as stochastic resonance~\cite{li2002, qiao2016, liu2019a}, 
driven escape dynamics~\cite{thompson1989, xie2003, sabbagh2026},
directed transports in terms of the ratchet effect~\cite{desouzasilva2006, angelani2011}, 
etc. However, the same has not been well explored methodically from a theoretical perspective for dynamic hysteresis. Some studies, revealing 
the effect of the asymmetry in the potential on this 
phenomenon, have been done for a variety of specific 
systems, such as energy harvesters~\cite{song2006, xuejun2013, wang2018, litak2022}, 
transformer circuits~\cite{huang1989, wang1992, moses2010, moses2011, aboura2016}, 
ferroelectric thin film materials~\cite{misirlioglu2010}, 
piezoelectric actuators~\cite{aguirre2012, litak2022}, 
electrochemical systems~\cite{gopan2020, vanderven2022}, 
etc. However, these studies are mostly experimental. Moreover, a few associated  numerical investigations that have been performed are based 
on the models which are specified for a particular system 
and do not include the effect of fluctuations. 
Therefore, from our current understanding, we assert that 
a general and complete theoretical picture is missing 
in the context of analyzing the role of asymmetry 
in the potential on the process of dynamic hysteresis 
in a systematic way. The implementations and the 
observations of the outcomes in this direction 
in a large number of systems of varied types, signify the 
importance of focusing on this particular aspect, which we 
consider for our present study.}\\

\mdas{We proceed by examining two distinct types of 
asymmetry in the potential. The first kind arises due to the 
different depths of the potential minima 
and corresponds to the energetic disparity between the two 
wells of the double-well potential. This energetic asymmetry 
is inherently observed in some physical systems, such as 
memristors, quantum dots, etc. mainly due to material 
inhomogeneity, defects, and impurities, and also in chemical 
environments corresponding to protein folding, molecular 
isomerization. The second variety results 
from the separate widths of the two potential wells 
and the asymmetric structure of the barrier separating them. 
This type can be classified as an entropic asymmetry which 
exists between the two wells of the system. 
We point out that despite their distinct forms, 
these two types of potentials can capture the  
basic features of the energetic and entropic asymmetric 
systems, respectively, in general. We consider that these 
are the two possible fundamental asymmetries that can 
exist in the governing potential. Therefore, the present
study with these two classes of systems mostly covers  
the generic effects of asymmetric potentials on dynamic 
hysteresis. It is worth mentioning that the first kind of 
the asymmetric potential, i.e., the energetic asymmetric 
case is more commonly regarded to understand the role of 
the underlying asymmetric potential on dynamical processes. 
Here, we also, importantly, take into account the second 
kind of asymmetric potential factor, arising as an entropic 
effect, to develop a complete understanding to address 
the present research question. Although the idea of this 
branch of asymmetry is comparatively new, 
its significance has been demonstrated in some recent 
studies~\cite{innerbichler2020, chupeau2020, rai2026}. 
Its presence is interpreted in physical devices; the particular examples of real systems where width asymmetries in the potential are observed include MEMS (Micro-Electro-Mechanical Systems), NEMS (Nanoelectromechanical systems), magnetic bistable systems with shape anisotropy, ferroelectric devices with defects, etc. The demonstration of the Brownian escape experiments in the optical tweezers setup with the width-asymmetric potential~\cite{chupeau2020}, and the existence of this characteristic asymmetry in various physical platforms, identify this class as one of the fundamental varieties. 
Therefore, we indicate that the 
consideration of this second kind of asymmetric 
potentials are significantly informative to establish 
an overall idea about the effect of the underlying asymmetry 
on the process of dynamic hysteresis. 
Overall, the realization of both types of asymmetric 
potentials across various existing systems indicates their 
importance in reality, apart from having purely theoretical 
significance. This provides the motivation to develop  
a systematic interpretation of their effects on the 
process of dynamic hysteresis, which often emerges in such 
bistable systems naturally or by design. 
The objective is to understand the control over the outcomes 
of dynamic hysteresis through the underlying asymmetry to 
benefit of practical applications.}\\ 


\mdas{A connected phenomenon to dynamic hysteresis is dynamic transitions. It is described in 
terms of the symmetry of the hysteresis loops around the origin~\cite{chakrabarti1999}. 
An order parameter is designated, which is, by definition, 
zero for the symmetric loops and has a non-zero value when the loops develop asymmetry around the 
span of the response function. The transformation between the symmetric phase, 
characterized by the zero value of the order parameter, and the asymmetric phase, defined by the 
non-zero order parameter, is described as dynamic transitions. For the classical case of 
symmetric systems, this transition to the asymmetric phase occurs under extreme conditions of the 
governing parameters, i.e., very low noise strengths, small amplitudes and large frequencies of 
the external oscillating field. This appearance of the asymmetric phase is the consequence of the 
fact that not all accessible states of the system are equally explored due to the limitations 
emerging from the range of the controlling parameter values associated with the process. 
For example, if the thermal fluctuation strength is too low, the dynamics get trapped in one of 
the two wells of the double-well potential. Therefore, the range of the response function, 
defined as the state variable or probability density in the two wells characterizing two states 
of the system, will not be covered symmetrically under the influence of the periodic force. 
Similarly, if the frequency of the external drive is very high, the system does not get 
sufficient time to relax and follow its intrinsic structure. 
This also leads to the asymmetry in the feedback, producing asymmetric hysteresis loops 
around the average value of the response function. In summary, for symmetric systems, 
the spontaneous symmetry breaking in hysteresis loops is observed in the limiting conditions 
of the control parameter values. In our present study, we examine dynamic transitions in 
asymmetric systems of both classes. Interestingly, we find that for the asymmetric setups, 
the asymmetric phases in the hysteresis loops (non-zero order parameter values) appear 
for moderate ranges of the guiding parameter values, as well. This finding highlights that 
the underlying potential has a fundamental role in determining the symmetry-asymmetry aspect of 
the response function, manifest in the form of the hysteresis loops. This is not just 
guided by the external periodic control and the thermal fluctuations, both of which do not 
produce any bias when averaged over a period. 
}\\ 


\mdas{The paper is organized as follows. The system and dynamics have been described in Sec. II. 
This section contains the introduction of the model for the process of dynamic hysteresis 
within the Langevin dynamics framework. Dynamics for both the energetic and entropic asymmetric 
setups and their dimensionless descriptions have been explained here. 
The numerical simulation results are presented and discussed in Sec. III. 
The significance of the response function and its time evolution are described to begin with. 
The appearance of the hysteresis loops is depicted, consequently. Next, the hysteresis loop, 
as a measure of the hysteretic effect, and the order parameter, as a quantifier of the 
dynamic transitions, are defined. Both of these central quantities of dynamic hysteresis and 
transitions are analyzed for the asymmetric systems with the varying degree of asymmetry in the 
potential. The findings provide an understanding of the quantitative control over these 
dynamic processes by the extent of asymmetry in the governing potential. All the results 
have been produced and discussed for the systems with both classes of asymmetry, 
energetic and entropic.The paper is concluded in Sec. IV}

\section{System and Dynamics}

\begin{figure*}[ht]
  \centering

    \centering
    \includegraphics[width=0.45\linewidth, height=5cm]{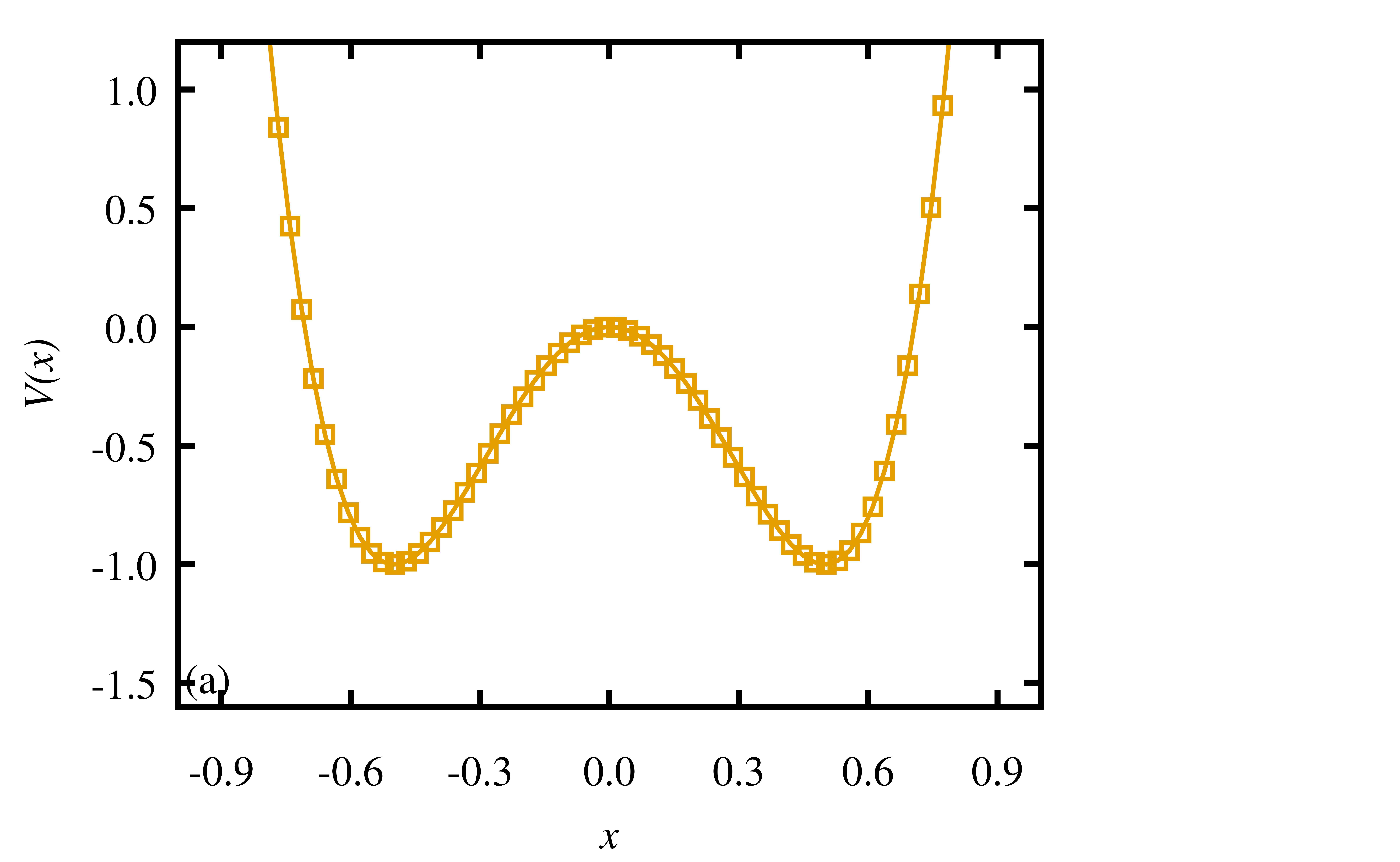}
    \includegraphics[width=0.45\linewidth, height=5cm]{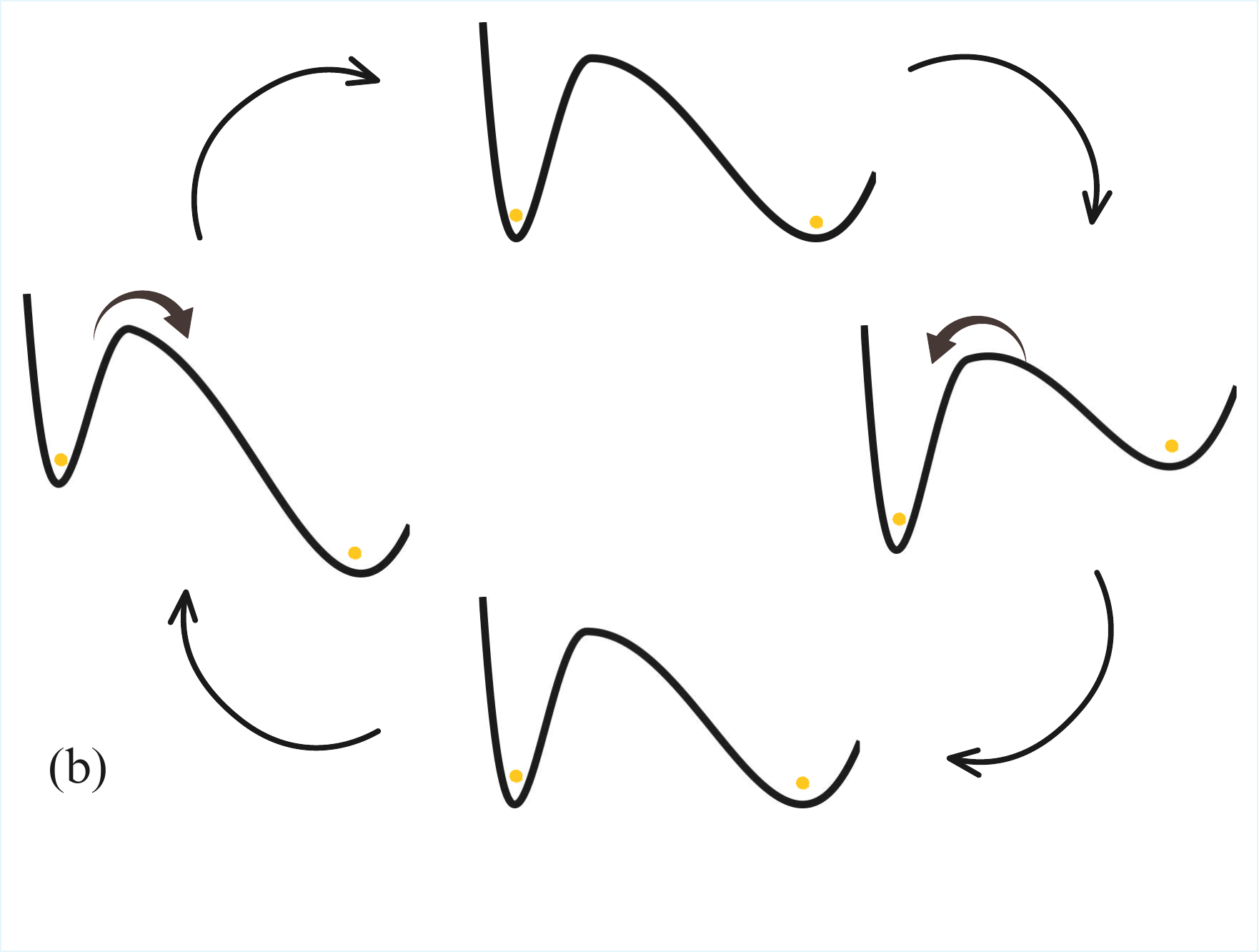}

\vspace{1.0cm}
  
    \centering
    \includegraphics[width=0.45\linewidth, height=5.5cm]{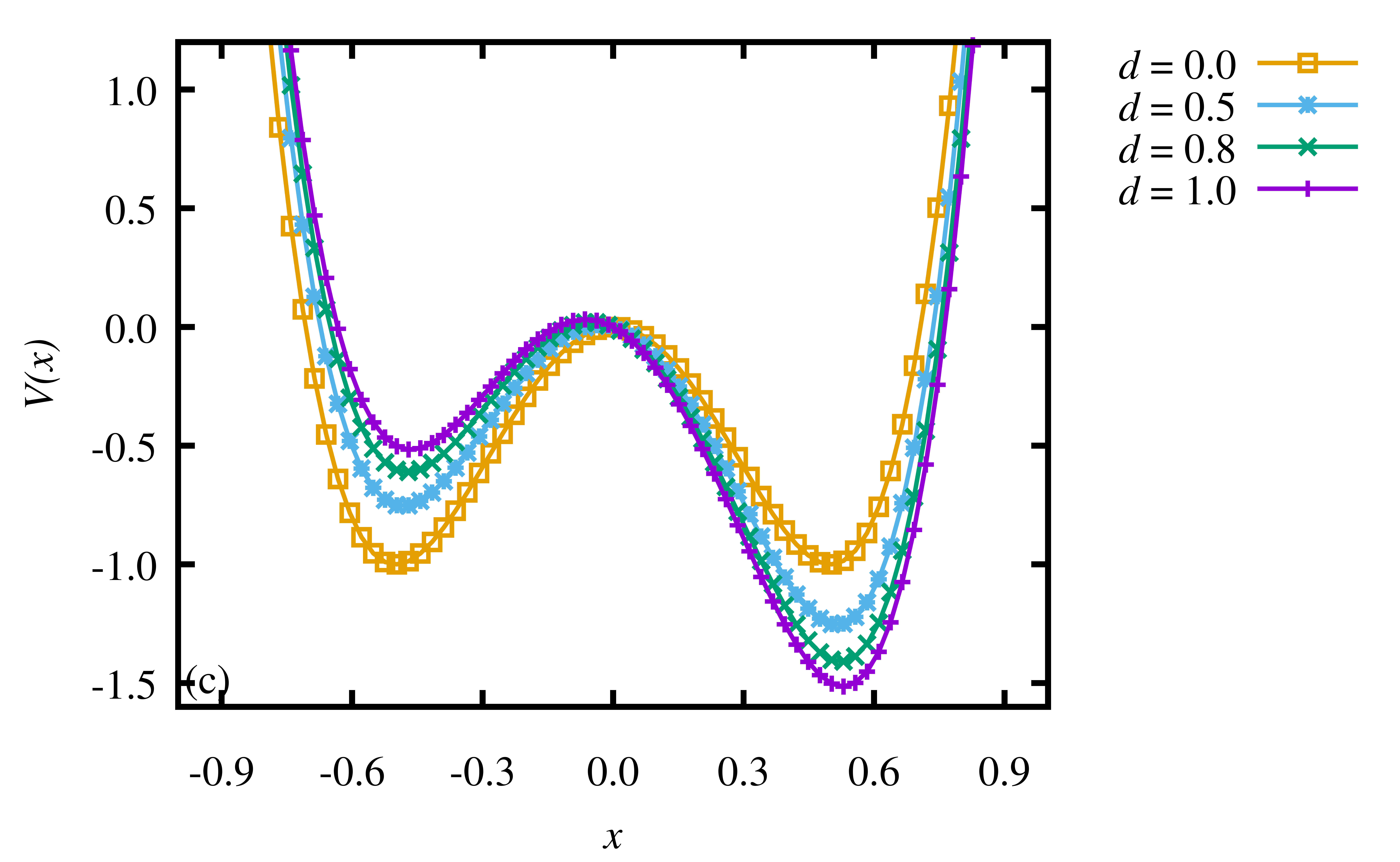}
    \includegraphics[width=0.45\linewidth, height=5.5cm]{Fig1d.pdf}

\vspace{1.0cm}

   \centering
   \includegraphics[width=0.45\linewidth, height=5.5cm]{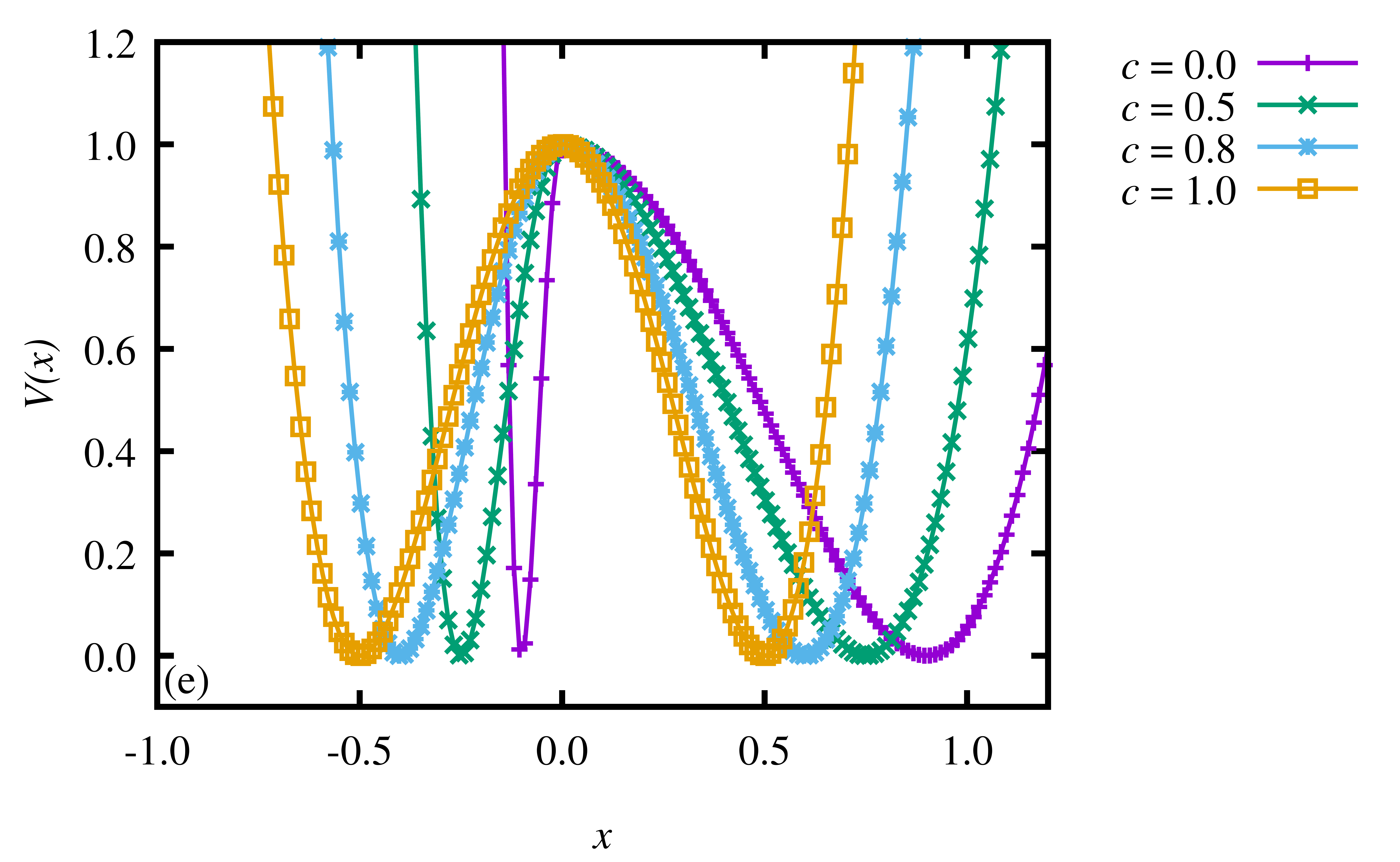}
   \includegraphics[width=0.45\linewidth, height=5.5cm]{Fig1f.pdf}

\caption{(a) Bistable potential in its symmetric form, (c) with different values of $d$ for the depth-asymmetric potential, and (e) with a range of values of $c$ for the width-asymmetric potential. Modulation in the shape of the (b) symmetric, (d) depth-asymmetric, and (f) width-asymmetric potential due to the external periodic force over a complete period of forcing.}
  \label{Fig1}
\end{figure*}

Dynamic hysteresis is modeled through the motion of a Brownian particle 
in a bistable or double-well potential subject to an external periodic force. 
The dynamics are represented by using the overdamped Langevin equation:
\begin{equation}\label{2.1}
    \gamma \frac{dx}{dt} = -\frac{\partial V(x)}{\partial x} + F(t) + \sqrt{D} \xi(t).
\end{equation}
Here, $x$ corresponds to the position of the particle at time $t$.  
\( \gamma \) denotes the friction or damping coefficient. 
\( \xi(t) \) is Gaussian white noise with statistical properties
\[
\langle \xi(t) \rangle = 0, \quad \langle \xi(t) \xi(t') \rangle = 2 \delta(t - t').\] 
It takes care of the environmental fluctuations, which play an important role 
in the process of dynamic hysteresis. 
Its amplitude is defined by the noise strength \( D \), 
which is connected to the temperature $T$ and the damping coefficient 
$\gamma$ through the relation \( D = \gamma k_{B}T \), 
where $k_{B}$ is the Boltzmann constant.

\mdas{Now, we discuss the significance of the potential function $V(x)$ controlling the dynamics. 
With the objective of analyzing the effect of the asymmetry in the potential 
on the current dynamical process, we consider two intrinsic types of asymmetries.  
The first one is associated with the difference in the depths of the two potential wells 
of the double-well potential and can be classified as producing an energy-asymmetric potential. 
The other type of asymmetry corresponds to the distinct widths of the two wells of the 
underlying potential. This creates asymmetry between the two wells of the potential, which is 
entropic in nature. The double-well potentials representing the bistable system 
for these two particular cases are modeled with two specified mathematical forms of $V(x)$, 
separately, to scrutinize two different situations. They both contain a tunable asymmetry 
parameter. However, there are unique differences in tuning for the two cases, which are  
implemented through the forms of the potentials.}

\mdas{For the former potential, $V(x)$ takes the form, 
\begin{align}\label{2.2a}
    V(x) = \dfrac{a}{4}x^{4}-\dfrac{b}{2}x^{2}-dx.
\end{align}
Here, the constant parameters $a$, $b$ and $d$ determine the 
barrier height and the positions of the minima of the potential. 
Particularly, the parameter $d$ introduces asymmetry in the 
energy levels or depths of the two minima of the double-well 
potential and controls the relative differences between their values.  
For the value of $d=0$, the potential takes a symmetric form.}

\mdas{For the second case, the potential form $V(x)$ is represented as, 
\begin{align}\label{2.2b}
    V(x) &= h \left[ 1 + \left( \frac{2x}{S(x)} \right)^4 - 2 \left( \frac{2x}{S(x)} \right)^2 \right],\\
\textrm{with} \nonumber \\    
    S(x) &= c\, \theta(-x) + (2 - c)\, \theta(x), \nonumber
\end{align}
Here, \( h \) is the measure of the barrier height. 
\( S(x) \) is a scaling function containing the Heaviside 
function $\Theta (x)$. This modulates the relative 
widths of the left and right wells, introducing asymmetry in 
the potential through the tunable parameter $c$. 
$c=1$ corresponds to the symmetric system.}

\mdas{At this point, we understand that the tuning parameter $d$ in the form of $V(x)$ 
in Eq.~(\ref{2.2a}) accounts for the depth asymmetry in the potential, 
whereas the tuning parameter $c$ in Eq.~(\ref{2.2b}) regulates the potential's width asymmetry. 
We indicate that the specific form of $V(x)$ considered above 
can represent systems with depth and width asymmetry, in general,  
in respective cases. The asymmetry exists between the two states of the system, i.e., 
the left well and the right well, in terms of energy and entropy, accordingly. 
The negative gradient of the potential 
$V(x)$ generates the intrinsic force field for the 
corresponding situations that appear in the dynamics.}

\mdas{The other force that influences the dynamics and is essential to study dynamic hysteresis,  
is $F(t)$.} This is an external periodic force which acts directly on the particle.  
It can have the form $F_{0} \textrm{sin}(\omega t)$ or $F_{0} \textrm{cos}(\omega t)$, 
$F_{0}$ being the amplitude and $\omega$ corresponding to the frequency of the force.\\


Now, the Langevin dynamics which we consider for the numerical simulations takes the specific 
forms described \mdas{by the following two equations for the depth and width-asymmetric cases, 
respectively.}
\begin{equation}\label{2.4a}
    \frac{dx}{dt} =(d+ bx -ax^{3} )+ F(t) + \sqrt{D}\xi(t),
\end{equation}

\begin{equation}\label{2.4b}
    \frac{dx}{dt} = 16 h \left\{ \frac{x}{S(x)^2} -4\frac{x^3}{S(x)^4}\right\} + F(t) + \sqrt{D}\xi(t).
\end{equation}

\mdas{Here, we mention that in the above equations (Eq.~(\ref{2.4a}) and Eq.~(\ref{2.4b})) 
all the quantities are dimensionless numbers. This convention has been followed as 
it offers a simplified description that becomes convenient for further analyses~\cite{dillenschneider2009, rai2026}.
We arrive at this form of the dynamics starting from Eq.~(\ref{2.1}), 
by choosing appropriate scaling factors for the quantities present in the 
original Langevin dynamics (Eq.~(\ref{2.1})). Here, we specify the scaling coefficients for the 
individual key entities present in the dynamical equation. 
The factor for the position variable $x$ is $l$, which is the distance between the two minima 
of the symmetric double-well potential. It has a numerical value equal to $1$. 
The same for the time $t$ is $t_{l}= \gamma l^{2}/k_{B}T_{R}$, here $T_{R}$ is a 
reference temperature. As a matter of fact, $t_{l}$ is twice the time taken by 
the Brownian particle to cover the average distance $l$. 
The scaling coefficient for the amplitude of the periodic force $F_{0}$ is $\gamma l/ t_{l}$. 
The mentioned quantities in the dynamics are made dimensionless when 
they are divided by the described coefficients. The frequency of the periodic drive $\omega$ 
becomes dimensionless when it is multiplied by $t_{l}$. In Eq.~(\ref{2.4a}) and Eq.~(\ref{2.4b}) 
$\xi(t)$ is the scaled noise term and $D$ is the scaled noise strength expressed by $T/T_{R}$.} 

\mdas{The first term within 
the parentheses in Eq.~(\ref{2.4a}) and the similar term multiplied by a factor of $h$ in 
Eq.~(\ref{2.4b}) represent the dimensionless forms of the intrinsic forces in the two cases. 
In Eq.~(\ref{2.4b}) $h$ denotes the dimensionless height of the potential barrier and 
$S(\tilde{x})$ is essentially a constant.} 
The forms of the potentials, \mdas{corresponding to these intrinsic forces, for the symmetric 
case, and} for different values of the asymmetry parameters \mdas{$d$} and $c$ have been 
presented in Figs.~\ref{Fig1}(a), (c), and (e). \mdas{The modulations of these 
elemental potential forms through the control of the external periodic force have also been 
illustrated in Figs.~\ref{Fig1}(b), (d), and (f).}\\
\\

\section{Results and Discussions}

\mdas{The key objective of the current study is to analyze the role of asymmetry 
in the governing potential on the process of dynamic hysteresis.} 
To interpret the mechanism of dynamic hysteresis and transitions, 
we need to analyze the response of the system towards the external periodic force.   
This can be done by defining appropriate response functions that can capture the effect of the 
external drive. 
\mdas{We understand that the driving force has a direct impact on $x$, 
the position or the state variable of the system. This is evident from the representation 
of the dynamics through the dynamical equations, Eqs.~(\ref{2.4a}) and ~(\ref{2.4b}).} 
Therefore, it can be inferred that $x$ or its functions can serve the 
purpose. However, the presence of fluctuations in the system 
induces stochasticity in the estimated value of 
the state variable $x(t)$ at a given time. 
In that case, its ensemble-averaged value, $\langle x(t) \rangle$, can 
be considered the reliable measure of the system's feedback. 
Another relevant and steady quantifier is the 
integrated probability of residence of the particles 
in either of the two wells of the bistable potential 
at a certain time. This is calculated for an ensemble of adequate size. 

\mdas{The quantity is defined as follows;} 
\begin{eqnarray}\label{3.1}
    P_{L(R)}(t) = \int_{x_1}^{x_2}P(x,t)dx.
\end{eqnarray}
Here, $P_{L}(t)$ and $P_{R}(t)$ are the normalized integrated probability 
of the residence of the particles in the left well and 
the right well at time $t$, respectively. 
\mdas{i.e., it implies the condition that 
$P_{L}(t)+P_{R}(t)=1$ at any instant of time.} 
Therefore, the values of $P_{L}(t)$ and $P_{R}(t)$ range 
from $0$ to $1$. A mathematical interpretation 
for this quantifier is that it is the integrated value of the 
probability of finding a particle at position $x$ at time $t$, $P(x,t)$, 
in the range of the state variable $x$, 
within the respective limits of $x_{1}$ and $x_{2}$.  
For the present case, $x_{1}$ and $x_{2}$ define 
the practical boundaries of the left and the right well, appropriately. 
\mdas{Ideally, the values of $x_{1}$ and $x_{2}$ are $-\infty$ and $0$, 
respectively, for the left well, and their corresponding values are $0$ and $+\infty$ for the 
right well. However, for the realistic simulation purposes, the $\pm \infty$ limits can be 
truncated to pertinent finite values beyond which the probability of finding the Brownian 
particle becomes negligibly small.}


\mdas{We consider the latter quantifier, i.e., 
the integrated probabilities, as a more effective 
response function for the current study. 
This is because it directly 
bears information about the two states of the system, 
the left well and the right well of the 
double-well potential representing the bistable system, 
in an integrated manner. 
It reduces the notion of a continuous double-well potential to a two-state model~\cite{zhou1990, mantegna1998, goulding2007}.
This can be considered as a general framework to study many natural and designed 
dynamical processes in which the transitions between these two states are the central theme. 
These phenomena range from escape dynamics~\cite{zhou1990, goulding2007}, 
the process of protein folding~\cite{zwanzig1997, wood2016} 
stochastic resonance~\cite{bulsara1991, mitaim1998, gudowska2004},
resonant activation~\cite{gudowska2004, das2013},
to chemical reactions~\cite{sinclair1965, waldherr2010},
and numerous others~\cite{mantegna1998}. 
It is true that the details of the behavior of the continuous state variable $x$ can become 
important in the study of some specific aspects of some given processes~\cite{bar1992, jia1996, mei2003}.
However, the overall populations of these two states, i.e., 
$P_{L(R)}(t)$ becomes significant to analyze several basic facets of 
the dynamical processes, which are mainly concerned 
with the transitions from one state to another. 
As the process of hysteresis can be seen as the system's ability to retain memory 
of its past state, we contemplate that it can be well analyzed within the scheme of 
the two-state model. For this setup, 
the necessary information related to the analyses of the response of the system 
can be provided by the integrated probabilities associated with the two states.} 

\mdas{We point out that $\langle x(t) \rangle$ can also be regarded 
as a potential response function to study dynamic hysteresis. However, from the forms of the 
asymmetric potentials (Fig.~\ref{Fig1}(c) and (e)), it is apparent that their spans are different 
for the distinct degrees of asymmetry. This is more prominent for the systems 
with width asymmetry (Fig.~\ref{Fig1}(e)). Therefore, the range of $\langle x(t) \rangle$ will vary depending on the 
values of the asymmetry parameter. Consequently, when we aim to understand the fundamental 
effect of the asymmetry of the governing potential on the process of dynamic hysteresis, 
the quantifier $\langle x(t) \rangle$ would introduce an additional quantitative dependence 
in the concerned measures of hysteresis for the specific asymmetry parameter values. 
On the other hand, for the entire range of the potential forms, $P_{L(R)}(t)$ will vary between 
$0$ and $1$. As a result, the actual control of dynamic hysteresis through the degree of 
asymmetry in the potential can be appropriately interpreted considering the 
integrated probabilities $P_{L(R)}(t)$.   
This is due to the fact that they do not impose extra quantitative characteristics 
on the calculations for individual asymmetric systems. In summary, this quantifier can deliver 
the essential information to fulfil the goals of the present study 
without introducing any subjectivity in the measurements for specified degrees of asymmetry. 
}




\mdas{Following the above justification, we proceed to investigate and compare 
the effect of depth and width asymmetry in the underlying potential on dynamic hysteresis, 
considering the quantifier $P_{L(R)}(t)$.} 
In practice, we calculate the quantifier $P_{L(R)}(t)$ numerically by simulating the 
dynamics of the ensemble of Brownian particles governed by the Langevin equations 
(Eq.~(\ref{2.4a}) and ~(\ref{2.4b})) using Heun's method. 
\mdas{The solutions of the first equation (Eq.~(\ref{2.4a})) provide the results 
for the depth-asymmetric cases. The data for the width-asymmetric conditions  
are obtained from the simulations of the second equation (Eq.~(\ref{2.4b}).} 

\mdas{During the simulations of the above two dynamical equations, 
the parameter values related to the potentials are chosen such that 
the positions of the two minima, the maximum and the barrier height of 
the associated symmetric potential remain the same for both cases. 
This condition is followed for the relevance of the comparison of the results 
in two different situations. Therefore, for the depth-asymmetric potential, 
the values of $a$ and $b$ are chosen to be $64$ and $16$, respectively. 
In the case of the width-asymmetric potential, 
$h$ is considered to have a magnitude of $1$. 
These choices of parameters lead to the 
fixation of the barrier height to be $1$ in both scenarios for the 
symmetric intrinsic potential forms. 
Moreover, the two minima and the maximum of the two symmetric structures 
appear at $\pm 0.5$ and $0$, respectively, which are obtained 
from two distinct mathematical forms (Eqs.~(\ref{2.2a}) and ~(\ref{2.2b})),  
for the current selection of the parameter values. 
In the case of the depth-asymmetric potential, the rising value of $d$ implies 
an increase in the extent of asymmetry, with the $d=0$ value corresponding to the symmetric 
structure. The maximum range of the variation of $d$ is from $0$ to $1.0$ 
in all sets of studies done. 
On the other hand, for the width-asymmetric potential, the smaller value of 
the asymmetry parameter $c$ implies a more asymmetric potential form, 
following from the mathematical expression of the potential. A value of $c=1$ represents 
the corresponding symmetric potential. The variation of $c$ is implemented within the maximum 
range of $1$ to $0.1$ for all studies related to the width-asymmetric potential.}

\mdas{Both the dynamics are simulated with a timestep of $\Delta t = 10^{-4}$.} 
The Gaussian white noise is generated using the Box-Muller algorithm. 
In most of our simulations, at the initial time, 
the particle is placed at the top of the barrier. 
Next, it moves into either the left well or the right well, 
depending on the resultant action of the periodic force and the noise. 
In the subsequent time steps, it is examined whether the particle 
is in the left or the right well. If the particle is found 
to be in the left well at any arbitrary time $t$, the case 
contributes to $P_{L}(t)$. For the reverse scenario, i.e., 
the existence of the particle in the right well at time $t$ 
adds to $P_{R}(t)$. The simulations and the mentioned 
examination are repeated 
for an ensemble of Brownian particles to find the fraction of the particles 
remaining in the left or the right well at the specified time points of observation, 
for the whole duration of the process. 
A total of $10^{5}$ number of trajectories is considered in the ensemble.  
For each case, the time evolution of $10^{3}$ cycles 
\mdas{of the external periodic force} is implemented. 
\mdas{The described method is performed with the mentioned details 
for both frameworks of depth and width-asymmetric systems 
in the context of the study of dynamic hysteresis.} 

\begin{figure*}[ht]

\centering

    \centering
    
    \includegraphics[width=0.49\linewidth, height=5.5cm]{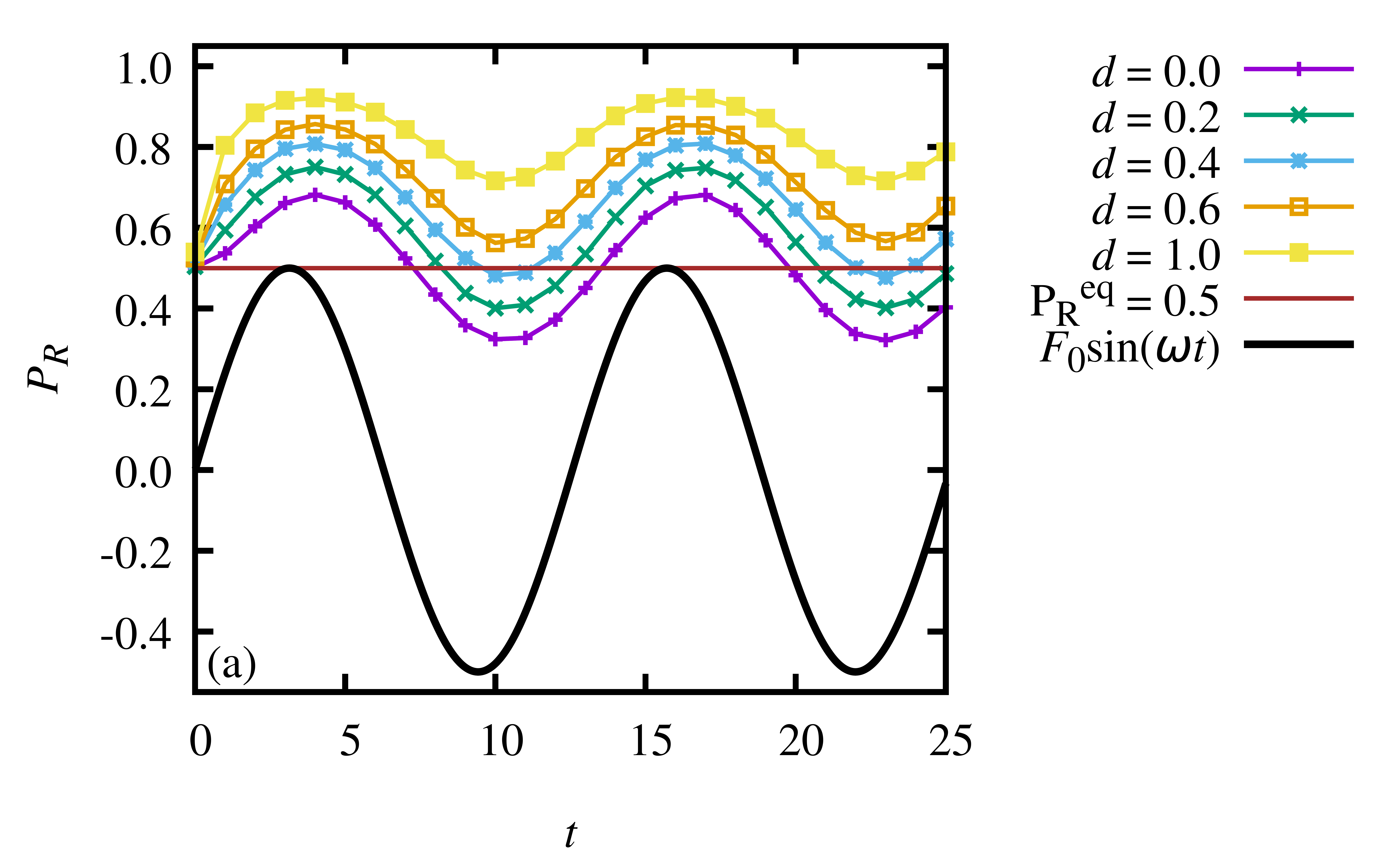}
    \includegraphics[width=0.49\linewidth, height=5.5cm]{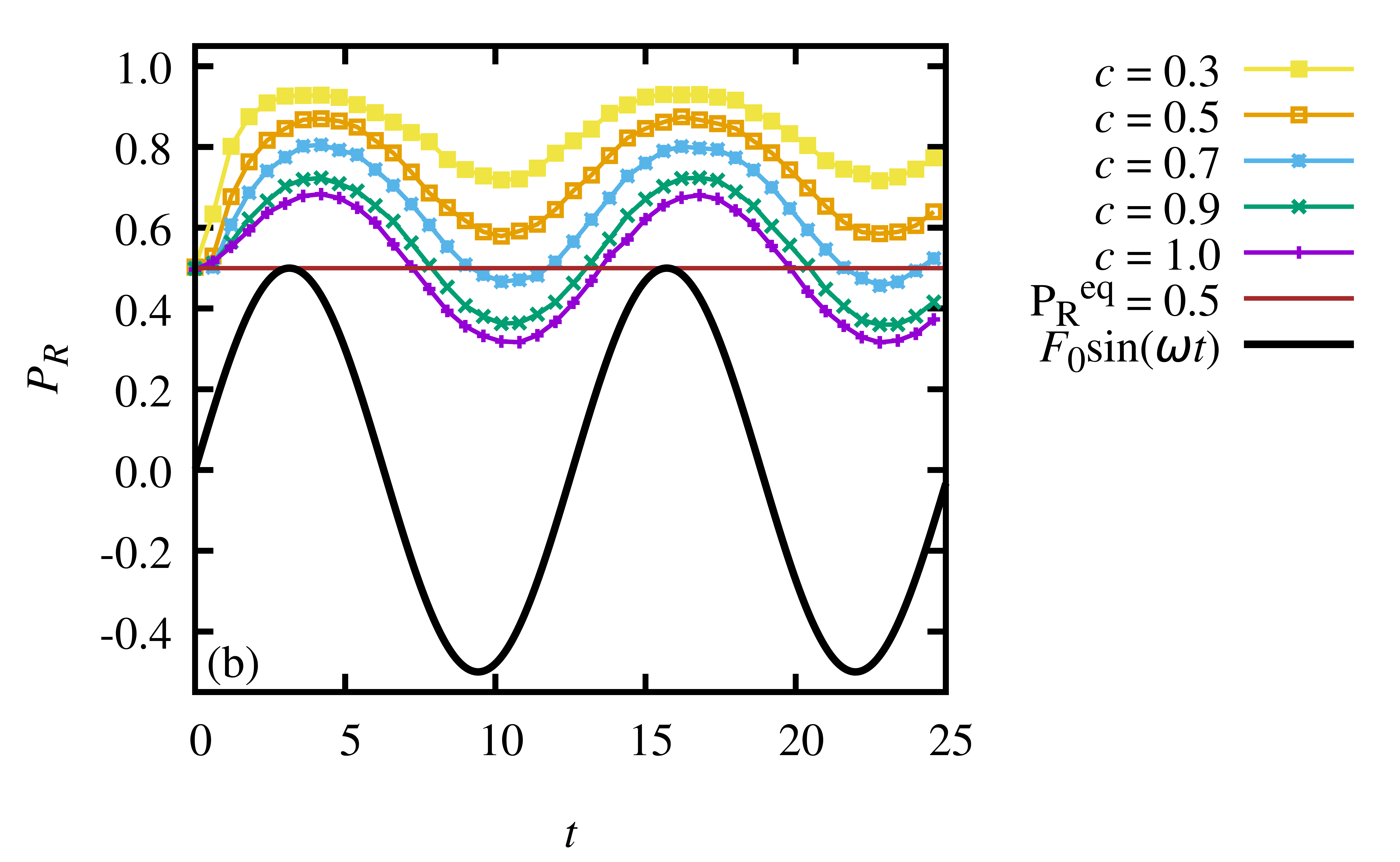}

  


    \caption{Lag: $P_{R}(t)$ vs $t$ in comparison to $F(t)$ vs $t$ at $F_0 = 0.5$, $D = 0.5$ and $\omega = 0.5$ for (a) the depth- or energetically asymmetric systems and (b) the width- or entropically asymmetric systems.\\ 
    }
    \label{Fig2a}
\end{figure*}

\mdas{To interpret the hysteretic effect in the systems under study, we start by comparing} 
the time evolution of the periodic force and the response function 
$P_{L(R)}(t)$ \mdas{over the periods} of the external drive. 
We analyze the above-mentioned variation for the 
symmetric potential and the potentials with different degrees of asymmetry. 
\mdas{We present the results for both kinds of asymmetric systems considered in our study.} 
We have particularly illustrated the comparative variation of a 
representative periodic force $F_{0}\textrm{sin}(\omega t)$ and $P_{R}(t)$ to 
directly understand the correspondence between the external periodic drive and the 
system's response towards it. \mdas{The results for two consecutive time periods 
of the external force are shown in Figs.~\ref{Fig2a}(a) and (b) 
for the depth and width-asymmetric systems, respectively, along with those 
for the corresponding symmetric setup.} 

We understand that, if we start all the trajectories 
of the ensemble at the top of the potential barrier, both $P_{L}(t)$ and $P_{R}(t)$ 
will be close to $0.5$, near the very beginning of the cycle. This is because the 
external force starts from zero and has negligible values near the starting time. 
Therefore, the particles feel almost no deterministic bias in this time domain.   
As a result, the occupancy of both the states in the left well and the right well becomes equally 
probable due to the effect of noise. 
Consequently, $P_{R}(t)$ in Figs.~\ref{Fig2a}(a) and (b) 
starts from $0.5$ for all cases studied. After that, it exhibits oscillatory behavior with 
respect to time. For the symmetric cases, we detect a clear lag in the response of 
the system. This is apparent as the maximum of the response function is shifted 
towards the right as compared to that of the periodic force along the time axis. 
\mdas{This lag persists for all asymmetric cases for both types of potential. 
Only the center of oscillation for $P_{R}(t)$ varies depending on the asymmetry parameter values  
$d$ (depth-asymmetric system) and $c$ (width-asymmetric system). 
In this particular observation, the $d$ range is varied between $0$ to $1.0$ 
and the parametric value of $c$ ranges from $1.0$ to $0.3$.}

For the symmetric potentials, the oscillation in $P_{R}(t)$ is observed around $0.5$, 
the equilibrium value of the observable without any external force. 
Whereas, as mentioned above, it oscillates around a different value, which is dependent on 
$d$ or $c$, for the asymmetric cases. 
This indicates a 
symmetry-breaking in the response. This happens even though the external drive does not induce any bias 
when averaged over a cycle. \mdas{We interpret that this 
symmetry breaking in the response emerges solely due to the 
asymmetry in the underlying potential, indicating its 
fundamental role in this dynamical process. 
Understanding the presence of lag in the 
response function with respect to the driving force for both 
categories of the systems under study, 
we proceed to perform thorough analyses of hysteresis in the 
characteristic setups with the relevant observables and 
measures related to the process of dynamic hysteresis.} 

\begin{figure*}[ht]

\centering

    \centering
    
    \includegraphics[width=0.49\linewidth, height=5.5cm]{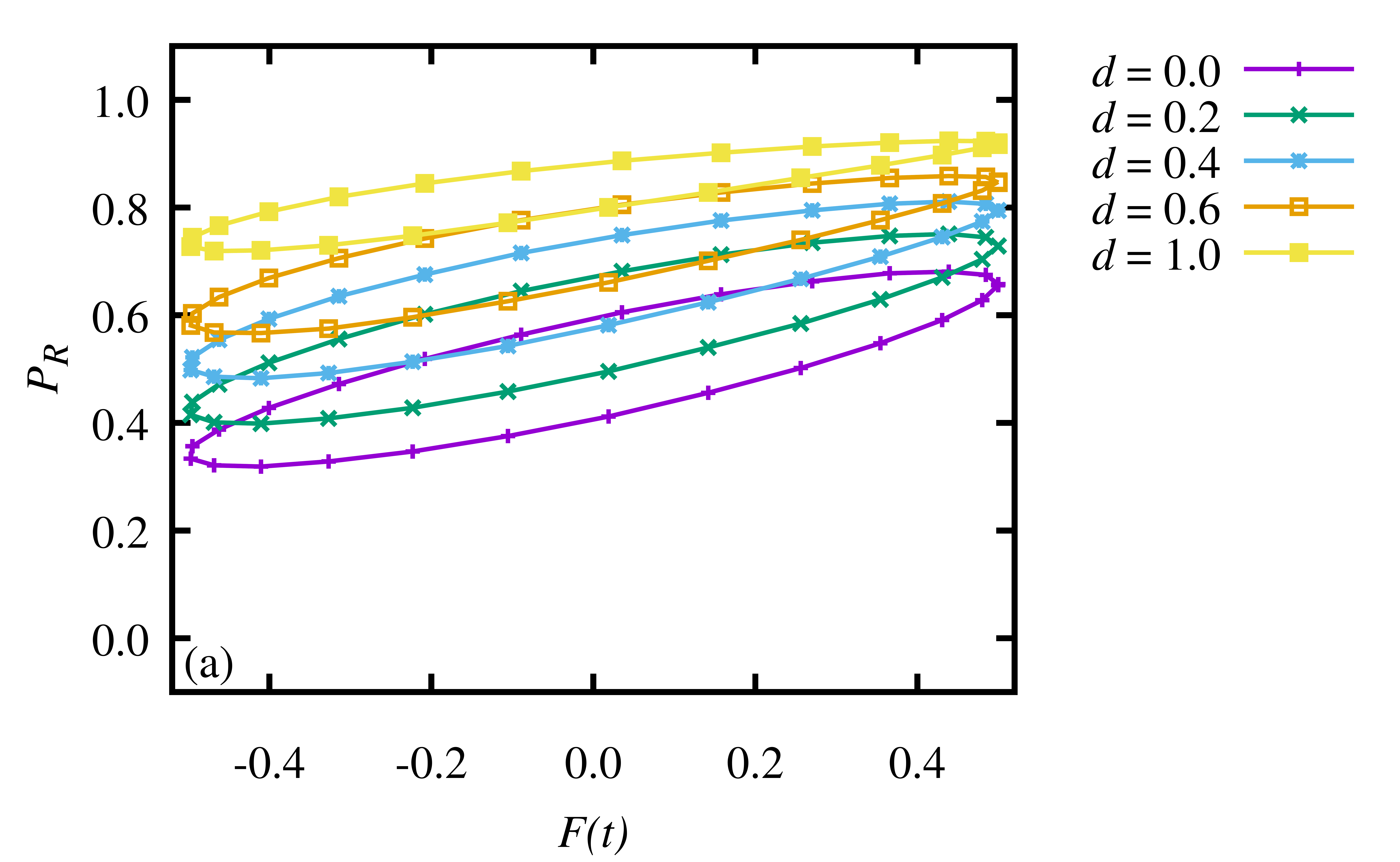}
    \includegraphics[width=0.49\linewidth, height=5.5cm]{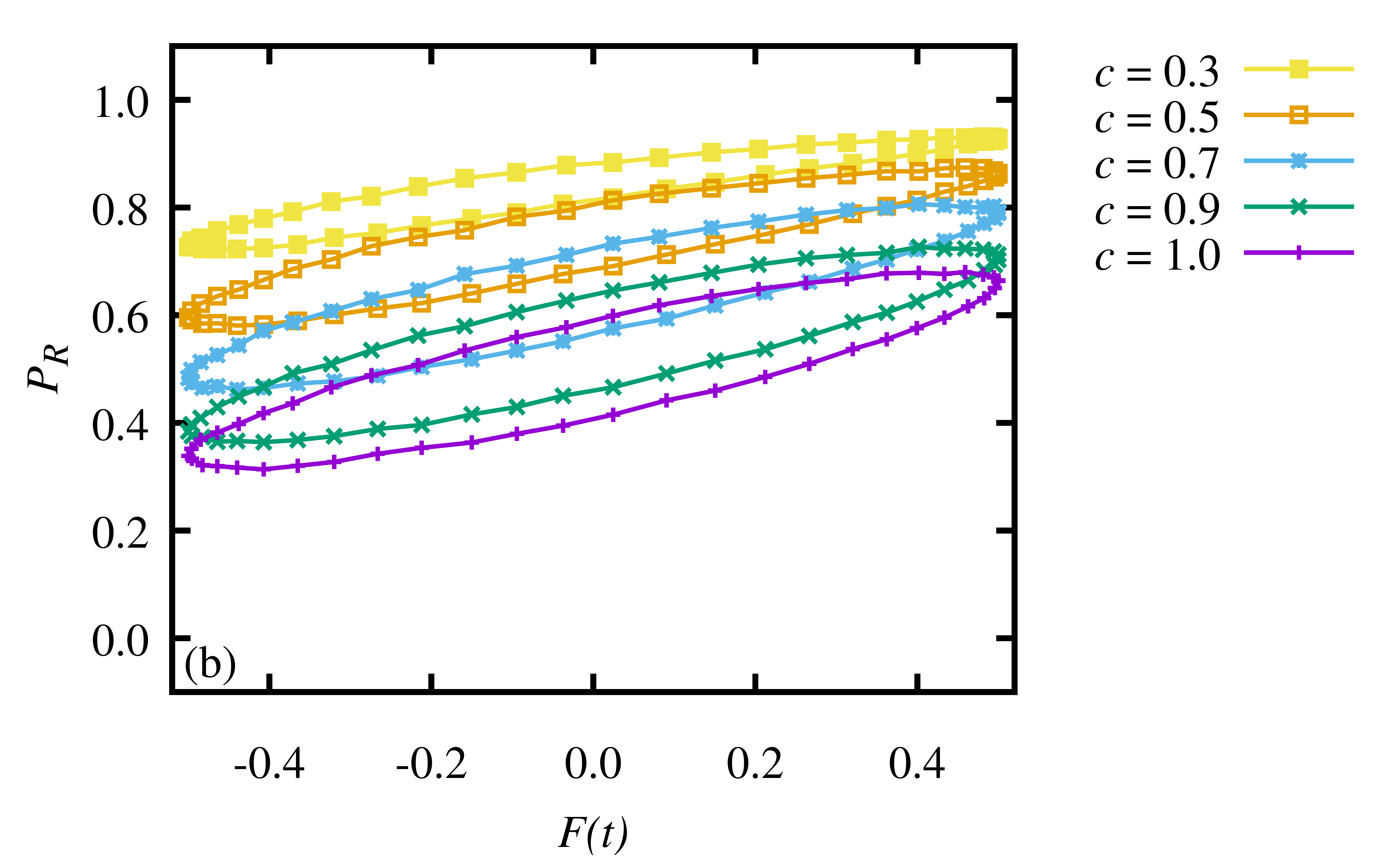}

  


    \caption{Hysteresis loops: $P_{R}(t)$ versus $F(t)$ at $F_0 = 0.5$, $D = 0.5$ and $\omega = 0.5$ for (a) the depth- or energetically asymmetric systems and (b) the width- or entropically asymmetric systems.\\ 
    }
    \label{Fig2b}
\end{figure*}
\mdas{To visualize the construction of the hysteresis loops and their features,  
we examine the variation of the response function $P_{L(R)}(t)$ 
as a function of the periodic force $F(t)$. 
The results have been presented with the $P_{R}(t)$ versus $F(t)$ plots 
in Figs.~\ref{Fig2b}(a) and (b) for the two classes of systems under study, 
for a range of asymmetry in each case.   
The formation of similar types of hysteresis loops is 
observed for both categories of systems. 
These are the steady hysteresis loops which 
are obtained after the initial transients are removed.} 

\mdas{
We have also scrutinized whether the natures of the hysteresis loops 
depend on the initial condition of the simulations. 
For this purpose, we consider three different starting conditions for the trajectories; 
one at the top of the barrier and two at the two distinct minima of the potential. 
The analysis suggests that for all different initial conditions, 
identically regular hysteresis loops, equivalent in terms of orientation and area,  
are formed after a few cycles. Only their initial transients differ based on the 
starting values of the position of the trajectories. This fact has been verified for 
both types of systems. The findings have been depicted with plots in the Supplemental Material 
(Sec. I).}

\mdas{The resulting hysteresis loops 
presented in Figs.~\ref{Fig2b}(a) and (b) are for an intermediate value of the 
driving frequency. 
To confirm the dynamic nature of the 
hysteresis studied in the present work, we also investigate the hysteresis loops for several 
values of the driving frequencies, ranging from very low to high. 
By definition, dynamic hysteresis occurs in the presence of the 
external periodic force and tends to vanish in the quasi-static limit. 
This is what is reflected through the almost-zero area hysteresis curves for very low 
values of the frequency of oscillation of the periodic 
drive. In the present study, it is observed for both 
symmetric and asymmetric systems of each kind. 
The graphical presentation of this description has been 
provided in the Supplemental Material (Sec. II). 
This particular observation of the vanishing loop area in the quasi-static limiting condition 
of the dynamical process validates the correctness of our numerical techniques. 
We make this point as our numerical simulations generate results 
that are expected and known at the specific limit for the symmetric setup.}

\mdas{The dynamic hysteresis loops as presented in Figs.~\ref{Fig2b}(a) and (b), obtained through 
our numerical approach suggest that they bear asymmetry for both types of 
asymmetric systems. Here, the non-symmetry in the hysteresis loops implies asymmetric 
behavior of the response function around its mean value, i.e., $0.5$, 
resulting in symmetry-breaking of the hysteresis loops. 
Symmetric hysteresis loops around $P_{R}=0.5$ are retrieved for the symmetric systems. 
We observe that the characteristics of the symmetry-breaking of the hysteresis loops 
are similar for the two classes of asymmetric systems under study. 
The extent of the deviation from the symmetric behavior of the hysteresis loops 
with respect to the change in the degree of asymmetry of the respective potentials, 
alters in the analogous manner for these two cases. This is evident from the temporal behavior 
of the response function (Figs.~\ref{Fig2a}(a) and (b)), 
and consequently from the hysteresis loops (Figs.~\ref{Fig2b}(a) and (b)). 
Therefore, we suggest that both types of asymmetry in the 
governing potential influence the behavior of the response 
function similarly during the process of dynamic 
hysteresis.} 

\mdas{We understand qualitative resemblance of asymmetry 
emerging in the forms of the hysteresis loops for the two types of 
asymmetric potentials. At this point, we further advance to carry out 
a thorough comparison between the two distinct categories in terms of 
a more quantitative estimate of the outcome of hysteresis, 
the hysteresis loop area, $A_{\textrm{hys}}$.}

It is known that the hysteresis loop area is an important measure of the hysteresis 
behavior exhibited by a system~\cite{chakrabarti1999,mahato1994}.
\mdas{This quantity signifies the area enclosed by the hysteresis curve.   
In other words, it is the area surrounded by the response function over a complete period of the  
driving force. Therefore, it is calculated through the cyclic integral of the response function 
over the evolution of the external drive over a full time-period~\cite{mahato1994, chakrabarti1999}. 
Consequently, in our case, it takes the following form,} 
\begin{eqnarray}\label{3.3}
A_{\textrm{hys}} = \oint P_{L(R)} (t) dF.
\end{eqnarray}

\mdas{Here, we aim to develop the quantitative understanding 
of the effect of asymmetry in the intrinsic potential on the process of dynamic hysteresis. 
Therefore, we observe the variation of $A_{\textrm{hys}}$ 
as a function of the asymmetry parameters $d$ and $c$, for two separate cases. 
We consider the change of $d$ within the range of $0$ and $1.0$. 
$c$ is varied between $1.0$ to $0.1$.}

\begin{figure*}[ht]

\centering

    
    \includegraphics[width=0.49\linewidth, height=5.5cm]{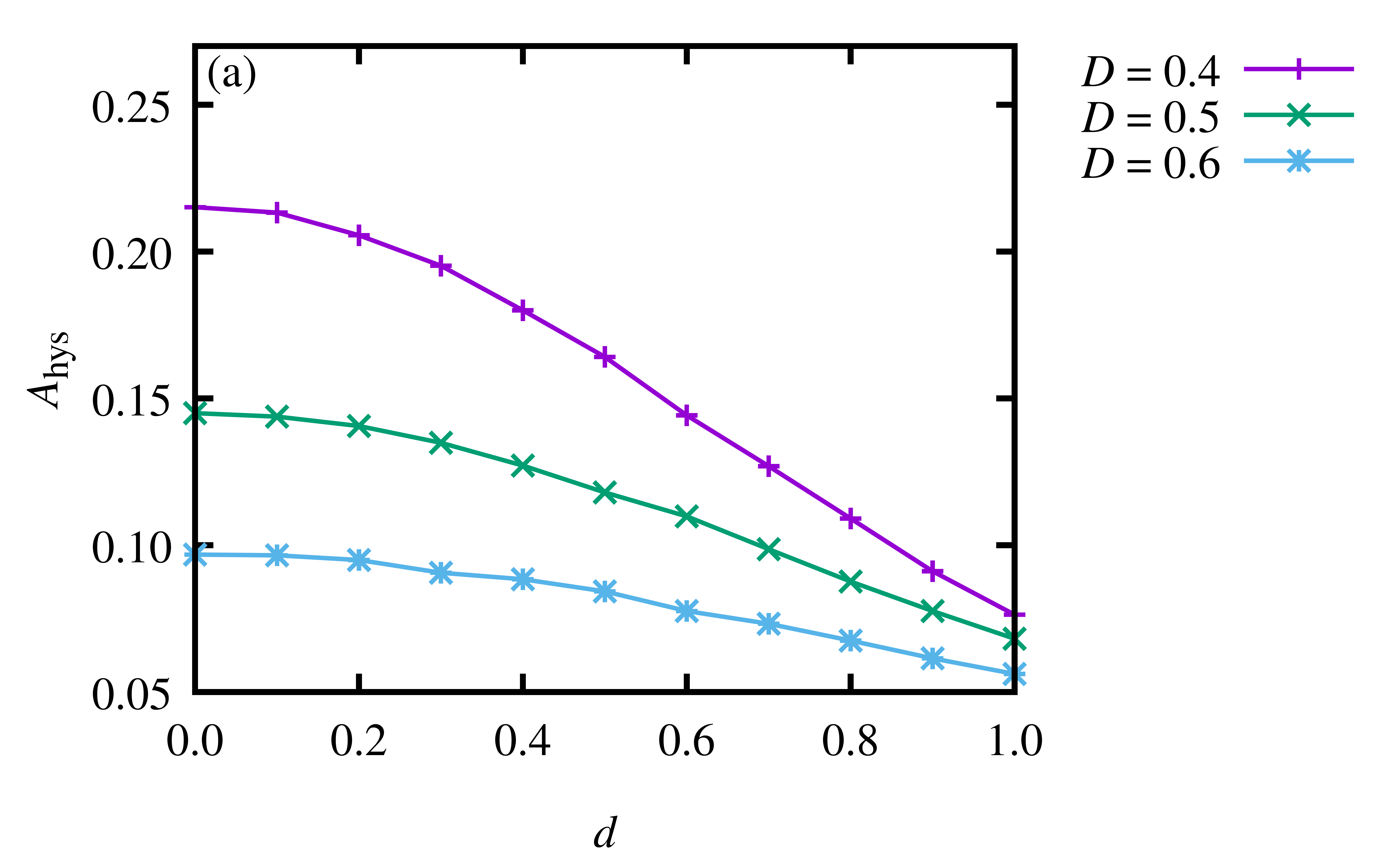}
    \includegraphics[width=0.49\linewidth, height=5.5cm]{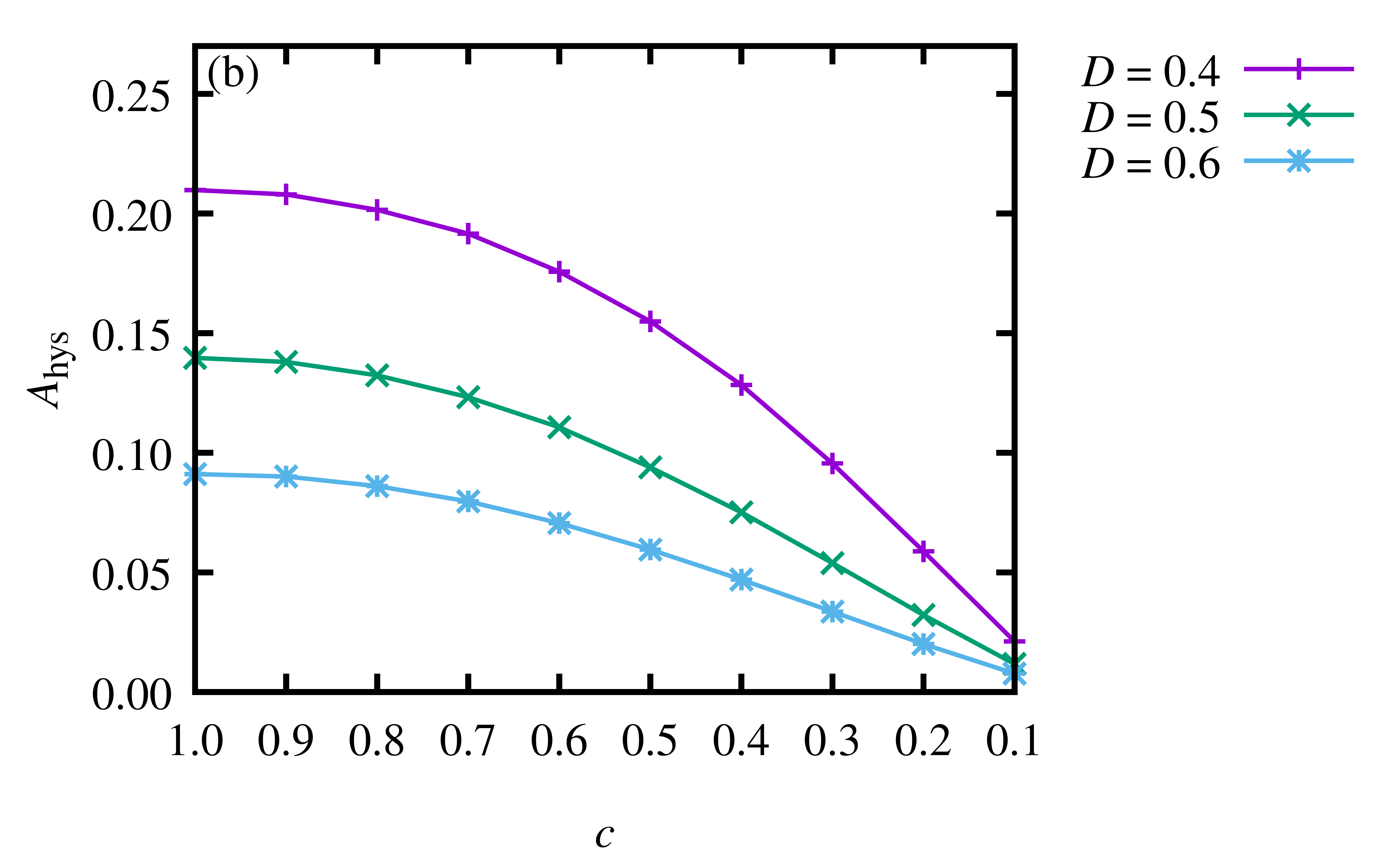}

\vspace{0.25cm}
  
    \centering

    \includegraphics[width=0.49\linewidth, height=5.5cm]{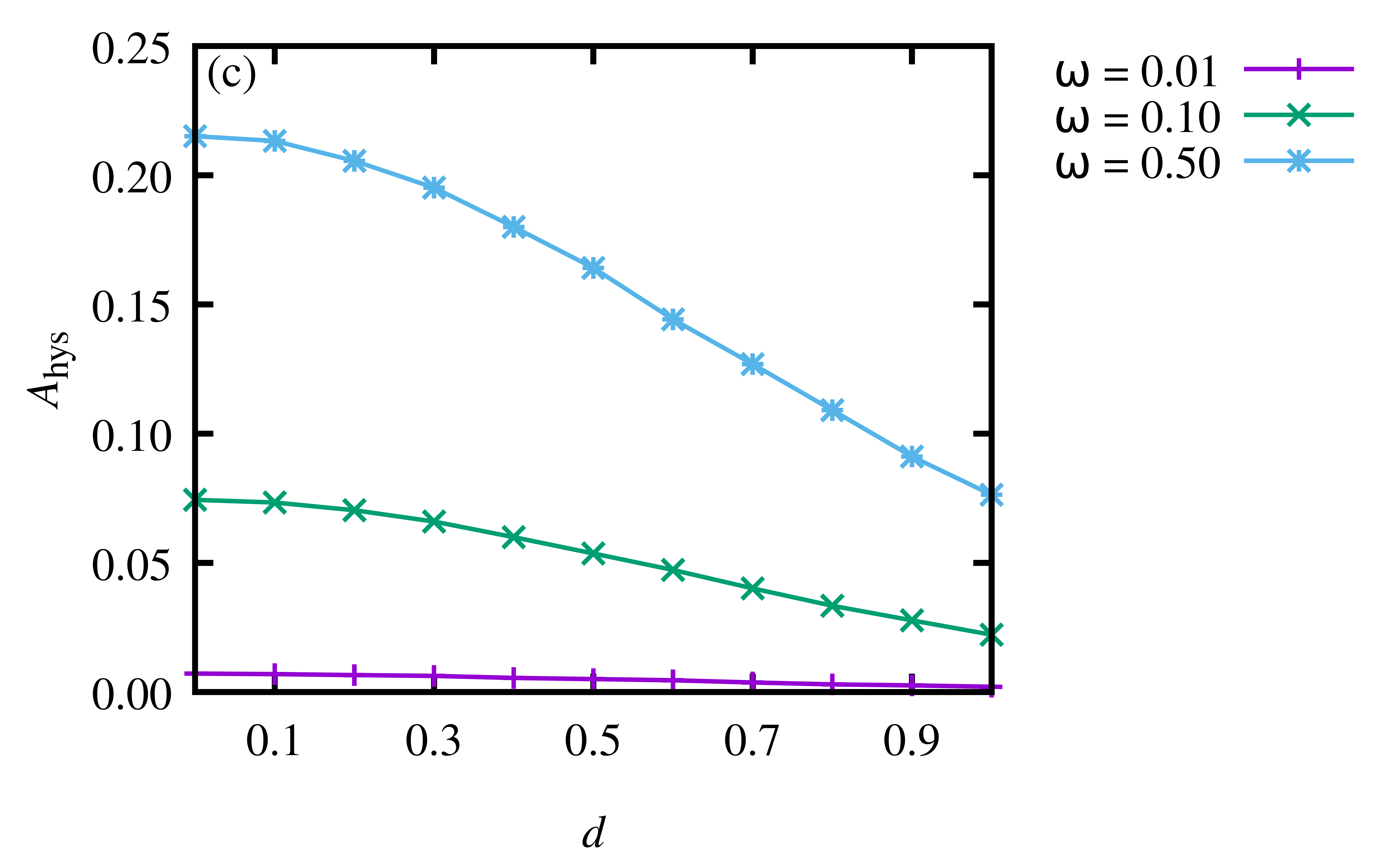}
    \includegraphics[width=0.49\linewidth, height=5.5cm]{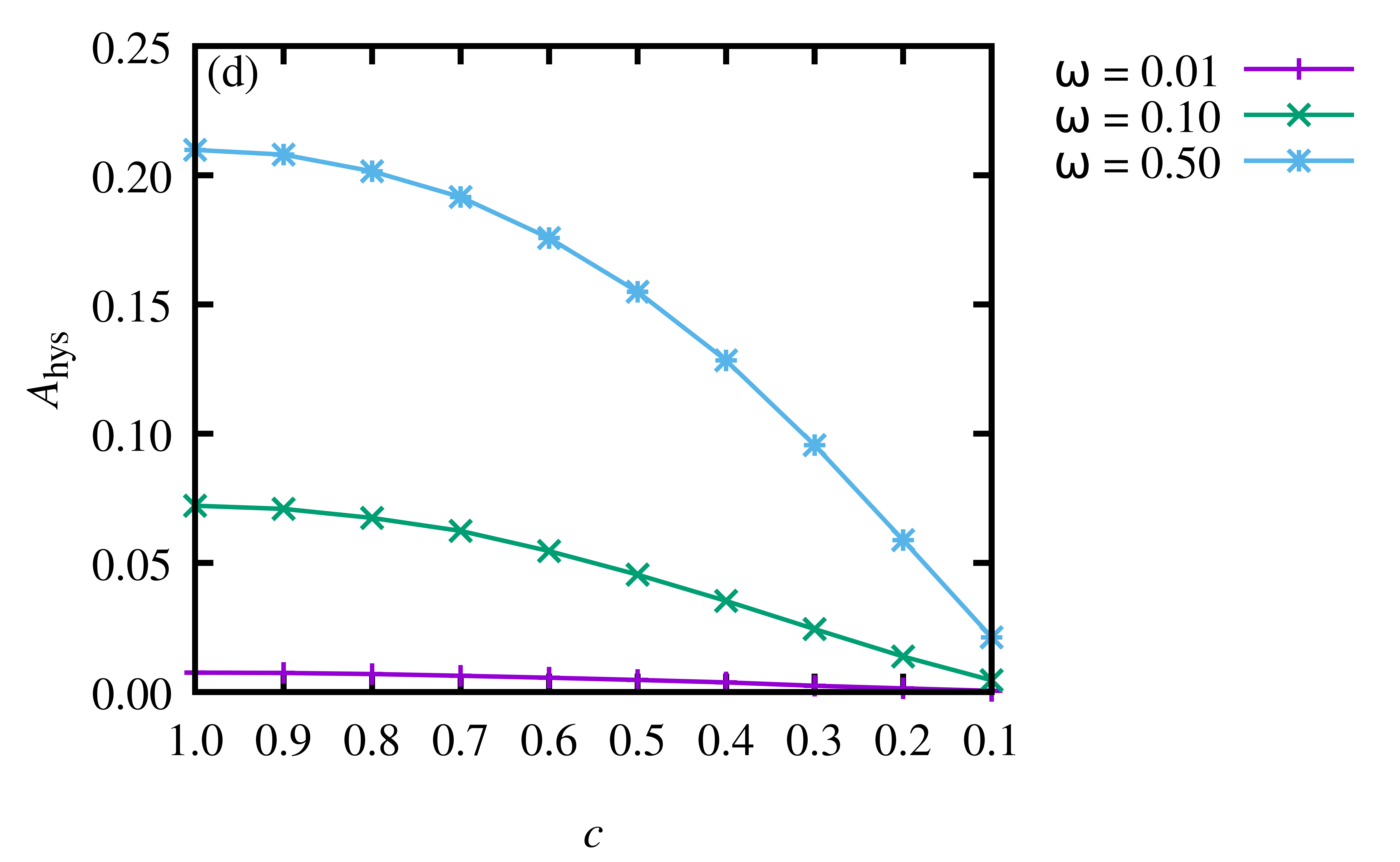}

    \caption{Hysteresis loop area versus (a) $d$ (for the systems with depth- or energetic asymmetry) and (b) $c$ (for the systems with width- or entropic asymmetry) at $F_0 = 0.5$ and $\omega=0.5$ for the parametric variation of $D$. Hysteresis loop area versus (c) $d$ (for the systems with depth- or energetic asymmetry) and (d) $c$ (for the systems with width- or entropic asymmetry) at $F_0 = 0.5$ and $D=0.4$ for the parametric variation of $\omega$.\\ 
    }

    \label{Fig2cd}
\end{figure*}

\mdas{For each category of the system, we studied two cases. 
First, we perform the analysis at a constant frequency of driving $\omega$, 
considering the parametric change of the noise strength $D$. 
The results have been presented in the Fig.~\ref{Fig2cd}(a) and (b) 
for the depth and the width-asymmetric potentials, respectively. 
Next, we carry out the study under the conditions of constant $D$, 
subject to the parametric change in $\omega$. 
The outcomes of these investigations have been illustrated in the Fig.~\ref{Fig2cd}(c) and (d) 
for the two types of potential forms. 
In these figures (Fig.~\ref{Fig2cd}(a) - (d)), along the $x$-axis, $d$ is plotted in the increasing order, whereas $c$ appears in decreasing magnitude. This has been done to ensure that the degree of asymmetry in the potential increases, for both types of systems, as one moves from left to right along the $x$-axis. 
In all cases, we detect that the hysteresis loop area has the maximum value for the 
symmetric potential under the present choice of parameters. 
It decreases monotonically with the increasing extent of asymmetry in the intrinsic potential. 
These are noticed for both classes of the system in all scenarios considered here.}

\mdas{The monotonic decrease of the hysteresis loop area with the increasing degree of asymmetry 
is the reflection of the reduced amplitude of oscillations of the response function. 
This implies that the external periodic force has less control over the alteration of the 
relative distributions in the two states for the asymmetric cases. 
In other words, the asymmetric systems are less responsive towards the extrinsic periodic perturbation, thereby reducing the impact of hysteresis. This also indicates that the systems with asymmetric structures cover a smaller span in the probability space when subject to periodic modulation. This effect changes steadily with the rising asymmetry in the underlying potential, resulting in a monotonic decrease in the hysteresis loop area 
with the increased extent of asymmetry.}

\mdas{We examine the existence of a scaling relation between $A_{\textrm{hys}}$ and the measure of asymmetry in the potential. We find that a scaling of the form $A_{\textrm{hys}} \sim c^{\gamma}$, with $\gamma \sim 1 $, holds for the system with entropic asymmetry for a moderate range of $D$ ($0.4-0.7$) and $\omega$ ($0.01-0.5$). 
We investigate a similar scaling relation for the system with energetic asymmetry, where we find the scaling to hold in the form $A_{\textrm{hys}} \sim (1-d)^{\gamma}$ for the same range of the parameter set as for the entropically asymmetric systems. In this case, the scaling exponent values seem to depend on the noise strength $D$. Here, considering $c$ and $(1-d)$ as the independent variables  certifies that we approach from the asymmetric to the symmetric systems for both cases. i.e., we scrutinize the scaling relations in similar scenarios for them. Some representative results for both categories of the systems  have been presented in the Supplemental Material (Sec. III).}

\mdas{The symmetry breaking that we observed in the dynamic hysteresis loops for 
the asymmetric potentials of both kinds can be analyzed within the purview of another 
fundamental phenomenon known as dynamic transitions~\cite{tome1990, sarkar1998, chakrabarti1999, korniss2000, marin2020, chattopadhyay2021}.}
The oscillation of the response function $P_{R}(t)$ around a different value apart 
from its equilibrium magnitude $0.5$ for the asymmetric cases leads to the emergence of 
this event for the current setups. 
This is understood in terms of the order parameter $Q$, which  
is defined for our present study as, 
\begin{eqnarray}\label{3.4}
  Q = \oint (P_{L(R)}(t)-P_{L(R)}^{\textrm{ref}}) dt. 
\end{eqnarray}
Here, $P_{L(R)}^{\textrm{ref}}$ has a value equal to $0.5$. 
\mdas{We designate this value as a reference point for $P_{L(R)}$ 
as the quantity oscillates around this mean when the systems are symmetric. 
This is the central point of oscillation of $P_{L(R)}(t)$ for the symmetric case, 
as the quantity attains this value of $0.5$ in the asymptotic limit 
in the absence of any external bias. 
When the systems become asymmetric, the centre of oscillation for $P_{L(R)}(t)$ 
differs from the reference value $0.5$ characterized by the extent of asymmetry in the system. 
As we try to understand the effect of asymmetry of the 
governing potential on this dynamical phenomenon under the influence of the periodic drive, 
we consider the symmetric case as the standard 
to quantify the asymmetry through the order parameter. 
The order parameter defined above gives a quantitative measure of the departure from the 
symmetric behavior of the response function during dynamic hysteresis 
due to the introduced asymmetry in the underlying potential.}

The dynamic transitions are interpreted by the 
distinctions of the zero and non-zero values of $Q$, which represent two unique dynamic phases. 
\mdas{$Q=0$ corresponds to the symmetric phase, signifying symmetric character in oscillation 
of the response function around its reference point during the process of dynamic hysteresis. 
The non-zero values of $Q$ are considered to form the asymmetric phase. 
It results from the asymmetry in the response of the system with 
respect to its standard reference value in the course of this dynamical phenomenon.}
In the classical case with a symmetric potential, 
the transition to the asymmetric phase, i.e., to the non-zero value of $Q$, is observed in the extreme 
conditions of the system and dynamics parameters, such as very low noise strength $D$, 
forcing amplitude $F_{0}$~\cite{tome1990, sarkar1998, chakrabarti1999, korniss2000}. 
However, interestingly, we observe that we can obtain non-zero $Q$ in moderate conditions of the above-mentioned quantities when the underlying potentials are asymmetric. 

\begin{figure*}[ht]

\centering

    \centering
    
    \includegraphics[width=0.32\linewidth, height=5.5cm]{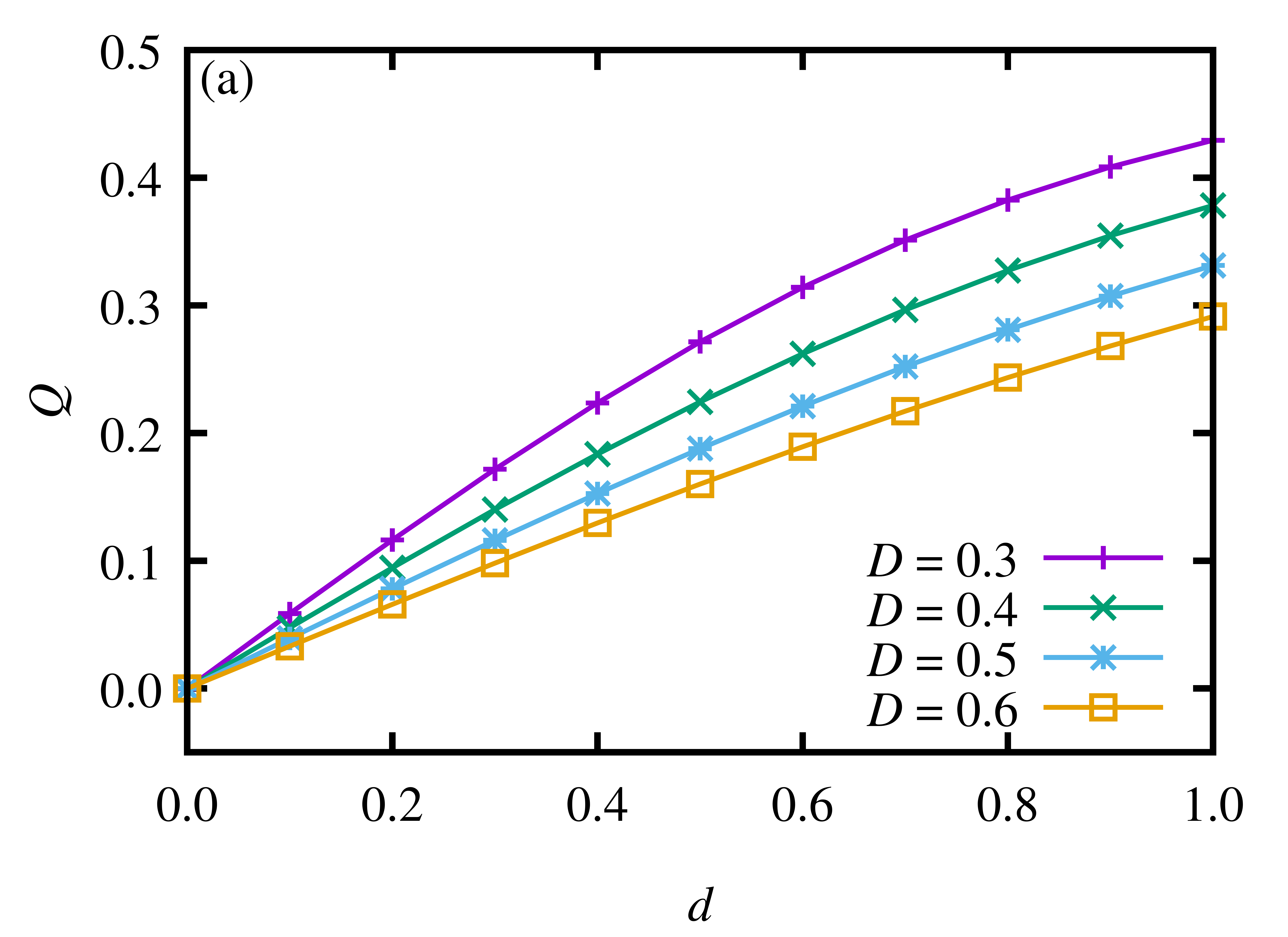}
    \includegraphics[width=0.32\linewidth, height=5.5cm]{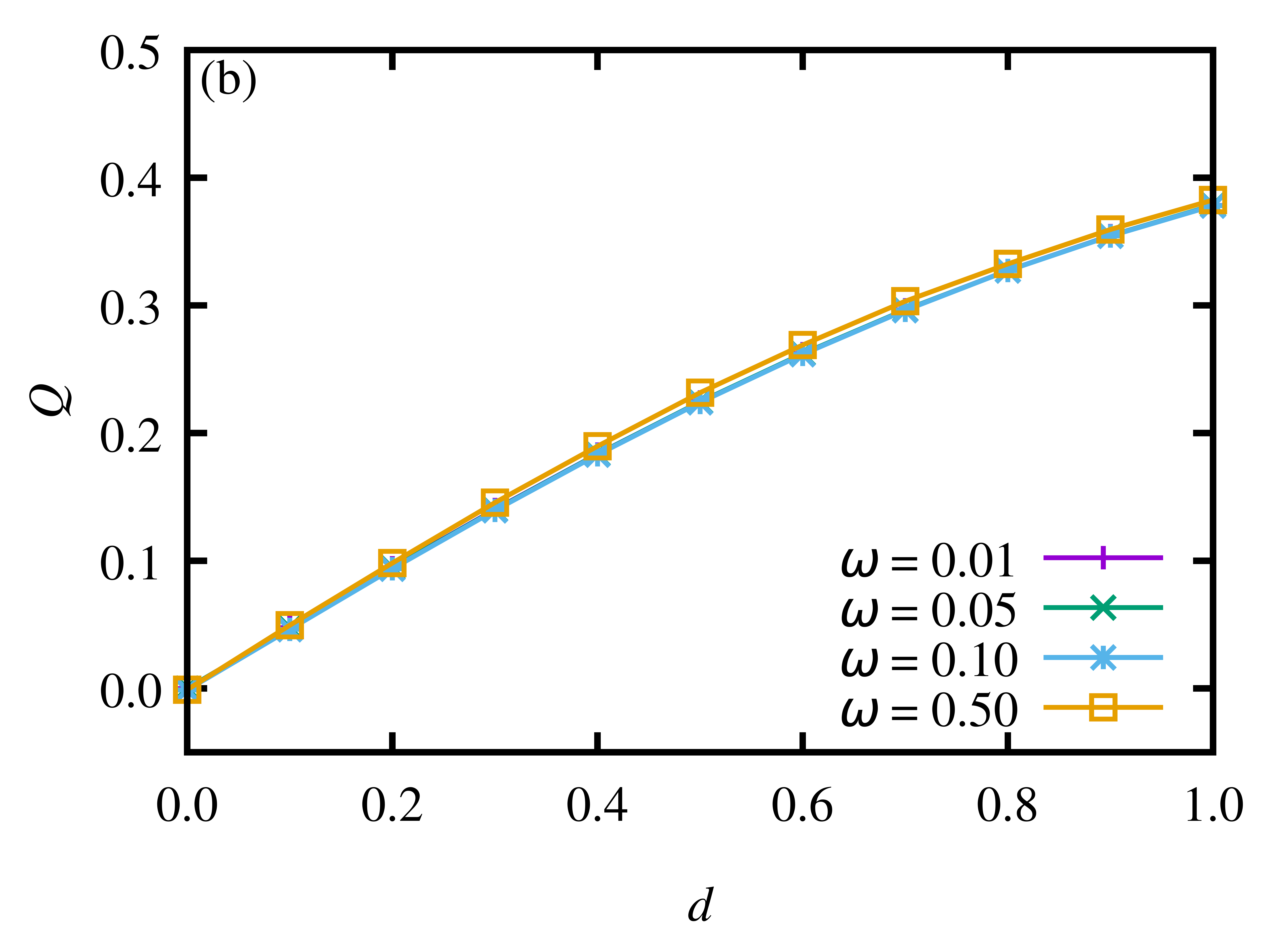}
    \includegraphics[width=0.32\linewidth, height=5.5cm]{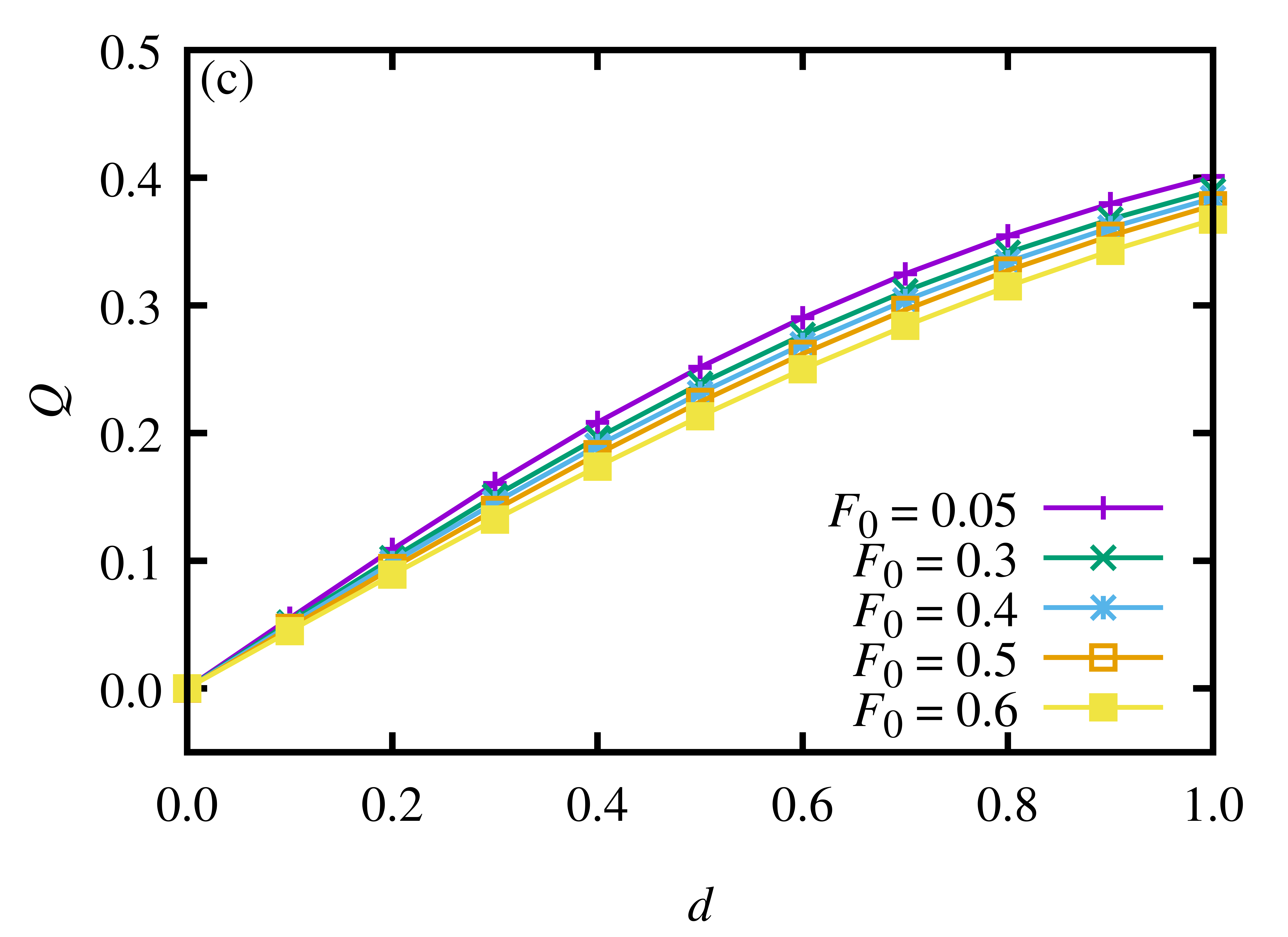}

    \centering
    
    \includegraphics[width=0.32\linewidth, height=5.5cm]{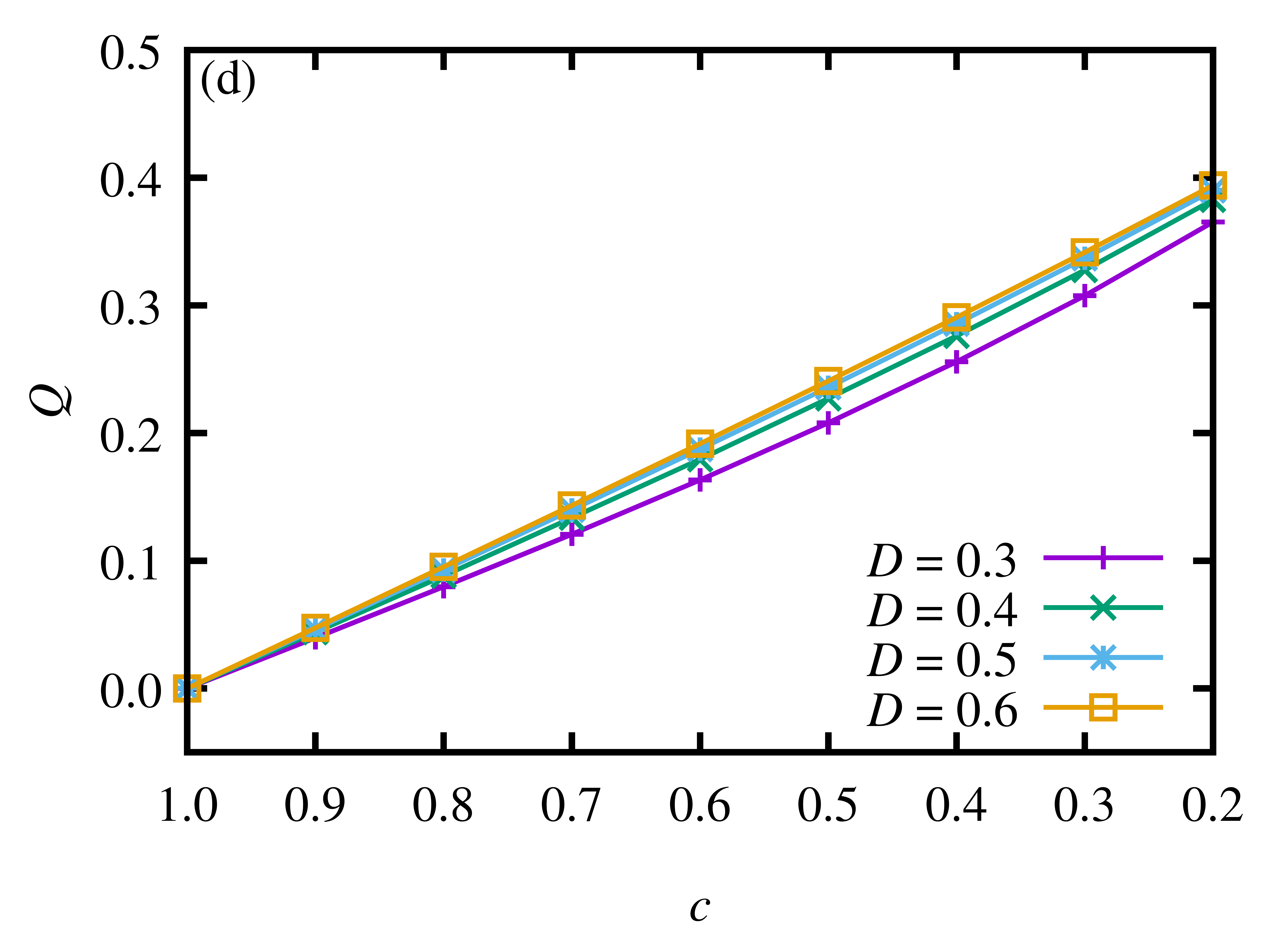}
    \includegraphics[width=0.32\linewidth, height=5.5cm]{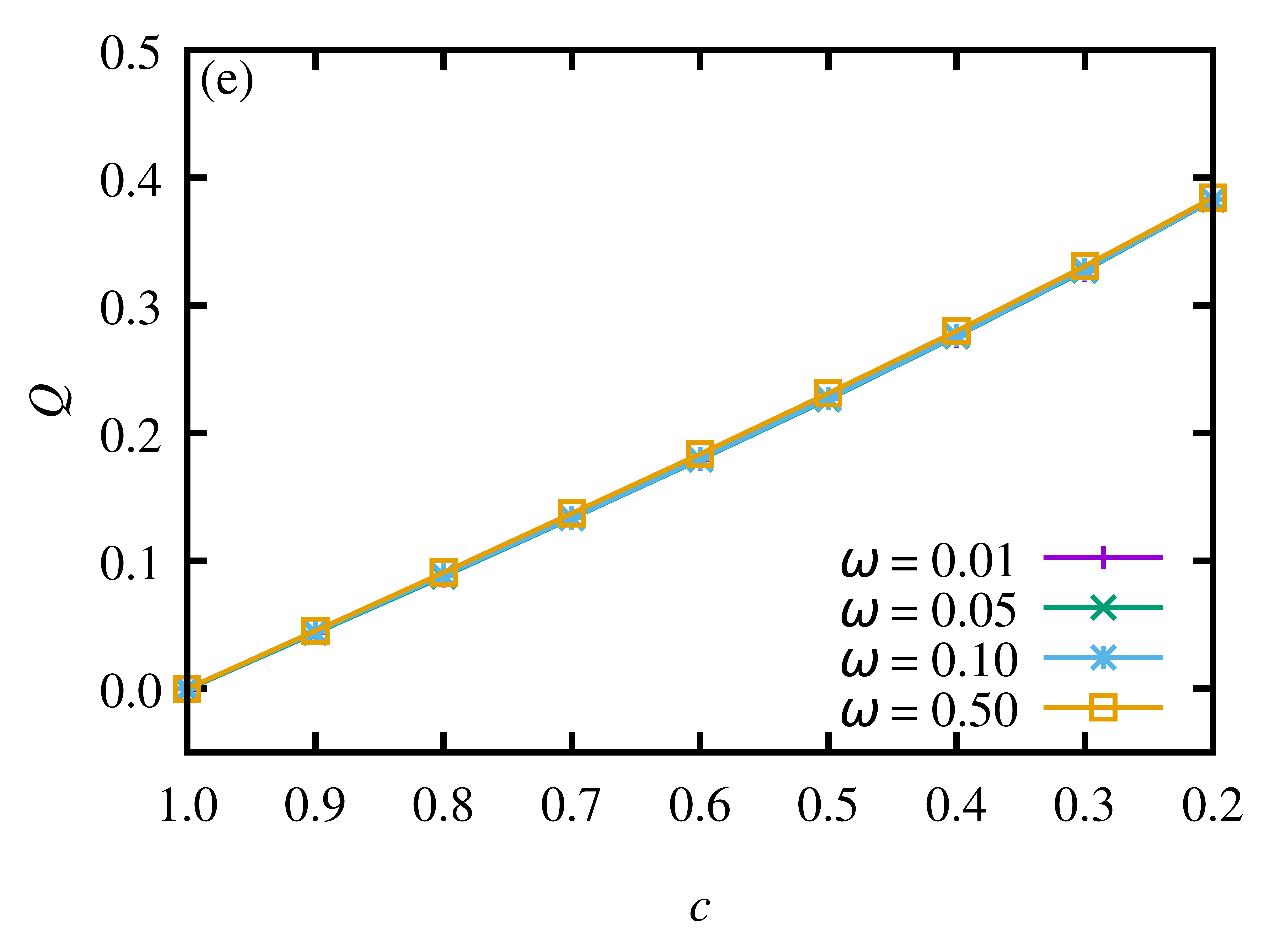}
    \includegraphics[width=0.32\linewidth, height=5.5cm]{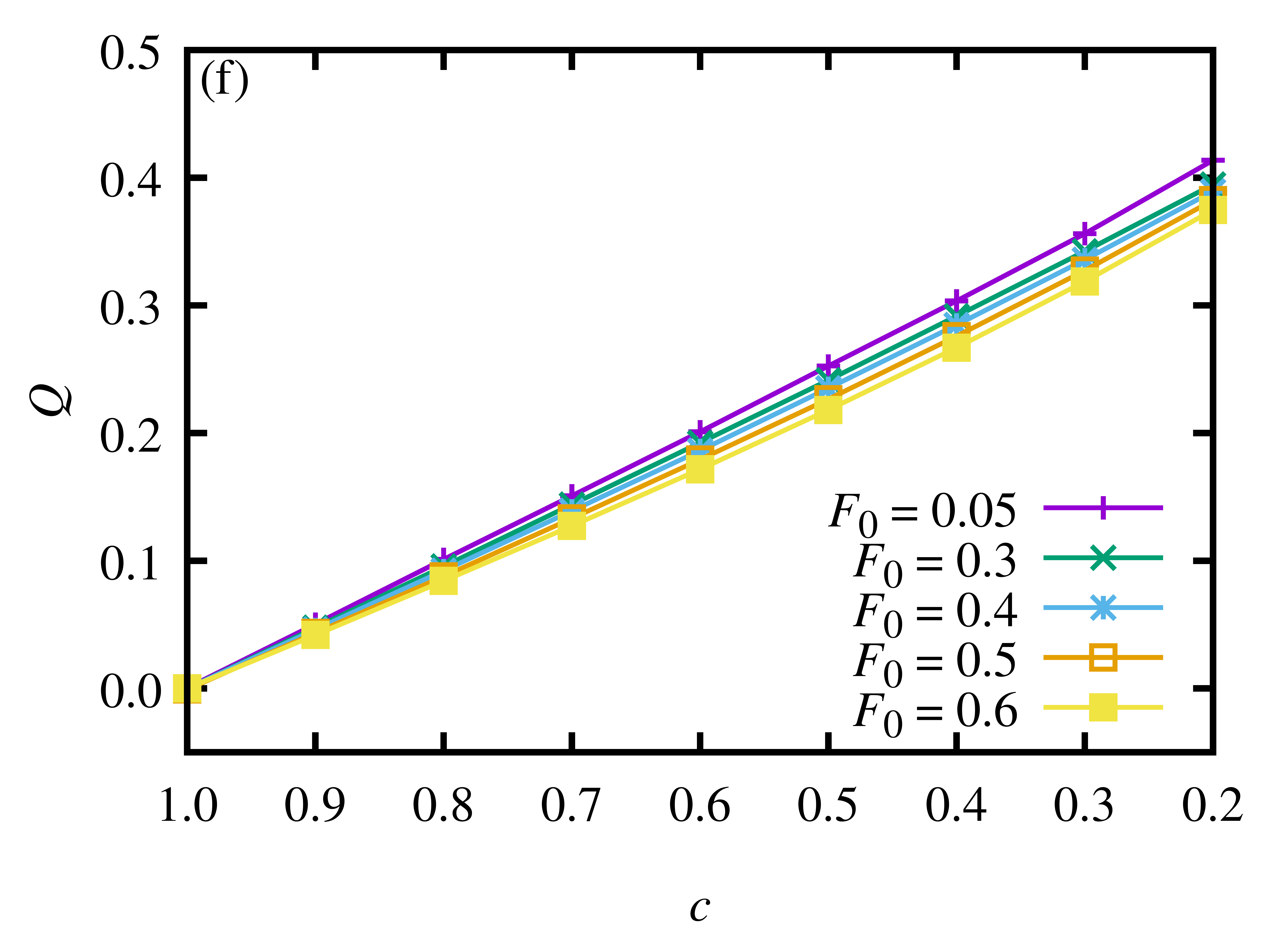}
    
    \caption{The order parameter versus (a) $d$ (for the systems with depth- or energetic asymmetry) and (d) $c$ (for the systems with width- or entropic asymmetry) at $F_0 = 0.5$ and $\omega=0.01$ for the parametric variation of $D$. The order parameter versus (b) $d$ (for the systems with depth- or energetic asymmetry) and (e) $c$ (for the systems with width- or entropic asymmetry) at $F_0 = 0.5$ and $D=0.4$ for the parametric variation of $\omega$. The order parameter versus (c) $d$ (for the systems with depth- or energetic asymmetry) and (f) $c$ (for the systems with width- or entropic asymmetry) at $\omega=0.05$ and $D=0.4$ for the parametric variation of $F_{0}$.\\  
    }
    \label{Fig3}
\end{figure*}

The transitions from the symmetric phase ($Q=0$) to the asymmetric phase (non-zero $Q$) 
have been shown in Fig.~\ref{Fig3}(a)-(c) \mdas{for the depth-asymmetric 
and in Fig.~\ref{Fig3}(d)-(f) for the width-asymmetric potentials. 
We observe the variation of $Q$ as a function of $d$ and $c$ in the respective cases. 
We notice that $Q$ starts from zero for the symmetric potentials for both categories 
of system. This is what is expected for the symmetric setups where hysteresis loops are 
symmetric around the reference value of the response function under moderate conditions. 
Then, with the rising extent of asymmetry in the systems, 
i.e., with increasing $d$ and decreasing $c$, the value of $Q$ increases. 
This is indicative of the enhancement of the degree of asymmetry 
with respect to the standard value of the response function 
in the dynamic hysteresis loops.}

\mdas{The above observations can be explained as follows. 
For the symmetric systems, the occupancies of both states, 
i.e., the left and the right wells are equally probable. 
Therefore, the external periodic drive modulates the extent of oscillation of the 
integrated probabilities symmetrically around the reference value $0.5$. 
Consequently, the cyclic integral calculated through Eq.~(\ref{3.4}) producing the 
value of the order parameter $Q$ becomes zero. 
On the other hand, the underlying asymmetry of the intrinsic potential 
creates a bias in the population of the two states of the system. 
This leads to the oscillations of the integrated probabilities around 
the mean value, which differs from the reference point $0.5$, 
generating non-zero values of $Q$ (Eq.~(\ref{3.4})). 
The extent of this departure of $P_{L(R)}^{\textrm{mean}}$ 
from the standard value of $0.5$ increases with the increasing degree of asymmetry in the system. 
As a result, $Q$ rises steadily for more asymmetric systems of both classes. 
For both frameworks, the studies have been done} for three different parametric variations 
with respect to the noise strength $D$ (Figs.~\ref{Fig3}(a) and (d)), 
frequency $\omega$ (Figs.~\ref{Fig3}(b) and (e)) 
and amplitude $F_{0}$ (Figs.~\ref{Fig3}(c) and (f)) of the external forcing. 
The results suggest that the quantitative nature of these 
transitions depends to some extent on $D$ and $F_{0}$; however, it remains almost 
invariant for the changing $\omega$. 
\mdas{These parametric variations have been analyzed in more detail, and the 
results have been presented in the Supplemental Material (Sec. IV).}

\mdas{The absence of the dependence of $Q$ on the frequency $\omega$ for a given degree of 
asymmetry for fixed $D$ and $F_{0}$ (Figs.~\ref{Fig3}(b) and (e)) suggests that 
the deviation of the mean of the response function from the symmetric reference 
point is similar for cycles of different time durations. This further signifies that 
the relative distributions in the two states of the system are not influenced 
by the frequency of the external control. This observation is the same for both  
systems with depth and width asymmetry.}

\mdas{The amplitude of the periodic force 
exhibits minor control over the order parameter $Q$ for the systems with the same extent of 
asymmetry (Figs.~\ref{Fig3}(c) and (f)). A larger value of the amplitude of the periodic force 
leads to a slightly lower value of the order parameter. This indicates that the greater amount of
force leads to a lesser relative difference between the occupancy of the two states of the 
asymmetric systems. Equivalently, it can be stated as a comparably smaller departure from the 
symmetric outcome of the phenomenon in the asymmetric setups under the action of the greater 
amount of the driving force. A justification of this finding can be explained as follows. 
The differences in the relative distributions in the two states, and hence the deviation from 
the symmetric behavior of the response function, 
are created by the introduced asymmetry in the potential. 
This effect is compensated by a larger magnitude of the periodic force.  
This is because it can act to spread the Brownian particles 
more uniformly in the two wells of the potential. 
The results are similar in this case also for the two types of asymmetric frameworks.} 

\mdas{We detect significant differences regarding the dependence of the variation of $Q$ as a function of the asymmetry parameter, on the noise strength $D$ 
for the asymmetric systems of two kinds (Figs.~\ref{Fig3}(a) and (d)). 
It is noticed that for a depth-asymmetric system with a given asymmetry parameter value, 
the noise strength $D$ has a substantial effect on the order parameter value $Q$. 
$Q$ decreases with increasing $D$ for a given depth-asymmetric system. 
Whereas $D$ does not impact $Q$ for a width-asymmetric system with a particular extent of 
asymmetry for which all other parameters have fixed values. 
This suggests that $D$ has a more direct effect in systems where energetic asymmetry is 
implemented. These observations can be interpreted thus. 
The higher value of $D$, signifying a larger temperature for the system,  
has the capacity to overcome the asymmetric behavior in the response 
caused by the energetic effect. This is reflected in the lower value of $Q$ for higher $D$ 
in the case of the depth-asymmetric systems. 
However, $D$ does not have much control over influencing the asymmetric response 
that arises due to the entropic effect. This is what emerges in the constancy of the 
value of $Q$ for a distinct width-asymmetric system when $D$ is varied.}

\mdas{Similar to the hysteresis loop area, we analyze the scaling relation of $Q$ with the extent of asymmetry for both types of systems. We find that the scaling relations of the form $Q \sim d^{\gamma}$ and $Q \sim (1-c)^{\gamma}$ hold for the systems with energetic and entropic asymmetry, respectively. The value of the scaling exponent $\gamma$ remains close to $1$ for both categories of the systems for the parameter set considered within the range of $D=0.4-0.7$ and $\omega=0.01-0.5$. Some representative scaling plots have been illustrated in the Supplementary Material (Sec. III). Here, the choice of the independent variable for scaling suggests that we move in the same direction in the symmetric-asymmetric aspect as the value of the independent variable rises for both classes of the systems.} 

\mdas{Collectively, these results suggest that the asymmetric feedback of the systems during dynamic hysteresis is primarily regulated by the asymmetry of the underlying potential for both types of asymmetric systems. The noise strength $D$ has a significant control over it of a secondary kind for systems with energetic asymmetry. This can be inferred from the dependence of the scaling exponent for the hysteresis loop area -- degree of asymmetry scaling relation on $D$ for the energetically asymmetric systems. The same fact is supported by the variability of $Q$ as a function of $D$ for the systems possessing the same extent of asymmetry for this class of systems. However, the sensitivity of the results on $D$ is not observed for the entropically asymmetric systems in similar scenarios. Therefore, these analyses imply that the temperature, through the strength of the thermal noise, influences the energetically asymmetric systems undergoing dynamic hysteresis and transitions to determine the outcomes of these processes. However, it does appear to affect the entropically asymmetric systems in this context. This aspect of temperature dependence in the present study can be explained as follows. For the energetic or depth asymmetric systems with different asymmetry parameter values, the barrier heights of the potentials change. However, they are the same for the systems with entropic or width asymmetry; in these cases, only the widths of the two wells differ for the systems with different degrees of asymmetry. Consequently, the barrier-height to noise strength ratio gets altered with the variation in the asymmetry parameter value and the temperature for the former class of systems. The barrier-height to noise strength ratio has significant effects on the consequences of the barrier crossing phenomena, such as dynamic hysteresis and transitions. We suggest this fact as the reason for the dependence of some results on the noise strength $D$ for the present study when the underlying potential possesses energetic asymmetry. Furthermore, the findings indicate that the driving frequency and the amplitude affect this asymmetric behavior in the response in similar ways for both categories of the potentials. The frequency does not seem to have any effect on it. The amplitude marginally modulates this asymmetric characteristic in the response of the asymmetric systems undergoing dynamic hysteresis and transitions.}\\

\mdas{Finally, we mention that the analyses and the interpretation described in the main manuscript are completely based on the numerical results. We suggest that a semi-analytical approach can be followed to develop a partial analytical framework for the current research question. This will involve the master equation for the integrated probabilities of residence and the modified Kramers' rate of transitions for the driven processes. We propose a scheme which produces results that show qualitative agreement with the findings obtained through the numerical simulation study. This has been outlined in the Supplemental Material (Sec. V). At this point, we suggest that the hypothesized semi-analytical method needs to be analyzed more thoroughly to propose it as an authentic analytical technique in the present context. We consider that the suggested semi-analytical model might be the starting point in this regard, as qualitative agreement with the numerical results is achieved. Further refinements of the scheme can generate a foolproof theory. It can be employed to get results for the systems undergoing dynamic hysteresis and transitions in the asymmetric setup with reduced computational cost. We consider that this important perspective can be explored comprehensively in future work.}  

\section{Conclusion}

\mdas{In the present study, we concentrate on the fundamental phenomena of dynamic hysteresis and 
transitions to understand the precise role of the underlying potential in controlling the processes. 
The effects on these mechanisms by the two essential components of the dynamics, the external 
periodic force; through its frequency and amplitude, and the environmental noise; quantified with
its strengths, are established. However, the aspect of the other basic factor, i.e., the structure of the intrinsic potential governing the dynamics, has not been investigated thoroughly in this connection. In the majority of the studies on dynamic hysteresis, the underlying potential has 
been considered to have a symmetric bistable structure, conventionally. Therefore, the influence of the asymmetry in the potential on this dynamical process has not been analyzed in detail. 
Some important experimental and numerical simulation studies in this regard have been carried out 
for specific systems individually. However, a general theoretical framework and, therefore, 
an integrated understanding of the definite effects of the asymmetry in the underlying potential on 
dynamic hysteresis and transitions was absent so far. Here, in this current work, we address this 
important point to develop a systematic interpretation of how the extent of asymmetry in the 
potential affects the dynamic hysteresis and transition phenomena.}

\mdas{The outcomes of our extensive numerical simulation study suggest that the hysteretic 
effect, quantified in terms of the area of the dynamic hysteresis loops, can be tuned by 
considering appropriate variation in the degree of asymmetry in the governing potential. 
Here, the asymmetry implies two different classes: 
one related to the asymmetric depths and the other to the asymmetric widths 
of the two wells of the model double-well potential implemented 
in the Langevin framework of dynamic hysteresis. 
It has been observed that for the wide range of parameter space scanned through our study, 
the dynamic hysteresis loop area primarily decreases with the increasing extent of asymmetry
for both types of potential systems. This fact indicates that the effect of hysteresis is maximum 
in the symmetric systems, and it diminishes steadily as the degree of asymmetry in the potential 
increases. 
We interpret that this effect emerges as a consequence of the fact that the external periodic 
control, which causes hysteresis, impacts the response function maximally for the 
symmetric systems. We suggest this through the examination of the time evolution 
of the response function, which shows maximum amplitude of oscillation for the symmetric systems. 
This amplitude decreases monotonically as the degree of asymmetry in the potential increases. 
This implies that the external drive cannot alter the response function, i.e., the probability 
distributions in the two states, which have been determined by the inherent asymmetric structure 
of the systems, significantly, for highly asymmetric systems. The smaller amplitudes of 
oscillations of the feedback function produce response function-field hysteresis loops of 
lesser width for asymmetric systems. Therefore, the reduced effect of hysteresis with the rising 
extent of asymmetry in the potential is reflected through the uniformly decreasing 
value of the hysteresis loop area.}

\mdas{One of the important directions of our study points out that it is possible to induce 
a dynamic transition to the asymmetric phase under moderate conditions in asymmetric systems. 
Here, the asymmetric phase refers to the hysteresis loops, which are asymmetric around the 
reference value of the response function. This appears even when the external periodic field does 
not provide any time-averaged bias to the system over a complete period. This observation is 
absent in symmetric systems under the same circumstances. In symmetric systems, 
symmetric hysteresis loops are observed for moderate parametric conditions. 
This analysis suggests that the underlying structure of governing potential is a prime 
decisive factor in determining the characteristics of the dynamic hysteresis and transition 
phenomena. It further establishes the significance of studying the symmetry-asymmetry aspect of the system in connection with these important dynamical processes. 
Similar to the methodical interpretation of the dependence of the dynamic hysteresis effect on 
the degree of asymmetry in the system, we develop a quantitative understanding related to dynamic 
transitions for the same. We examine the variation of the order parameter, quantifying the extent 
of asymmetry in the hysteresis loops, as a function of the asymmetry parameters of the 
potentials. The results suggest that the broken symmetric phase or the asymmetric phase appears in 
all asymmetric systems. Moreover, the extent of asymmetry in the hysteresis loops increases 
regularly with the rising degree of asymmetry in the intrinsic potential.}\\

\mdas{At this point, we assert that the two fundamental types of asymmetry in the potential, 
the depth and the width asymmetry, have similar effects on the measures of dynamic hysteresis 
and transition. This is understood through the analyses of the loop area for dynamic hysteresis 
and the order parameter for dynamic transitions. The results suggest that the dynamic hysteresis 
becomes less effective and the dynamic transitions become more pronounced as we increase the 
extent of asymmetry in the system. It shows that similar control over the outcomes of these 
dynamic processes can be achieved by introducing either class of asymmetry in the system. 
Also, if the systems' inherent asymmetries are of distinct types, one can expect to observe 
analogous consequences on these mechanisms with a varying degree of asymmetry for both types of 
systems. However, closer analyses indicate that the temperature appearing in 
the strength of the environmental fluctuations often has notable effects 
on the dynamical responses for the depth-asymmetric systems for the
processes considered here. Whereas it is not present for the width-asymmetric systems. This suggests that the depth-asymmetric systems are more responsive towards temperature variation in the current context.}

\mdas{In summary, our present work establishes a general model through which the important role 
of the underlying potential can be studied thoroughly for the process of dynamic hysteresis and 
transition. The numerical simulation study within this Langevin dynamics framework gives 
substantial scope for systematic understanding of the control through the extent of asymmetry 
in the system on these dynamic processes. We understand that the regulation of the outcomes 
of dynamic hysteresis and transitions has immense applicability to generate improved functions of 
devices of many kinds. Therefore, the idea regarding tuning these effects through the methodical  
control over the potentials, developed through the current study, is believed to contribute 
significantly to this domain of research. We investigate two classes of asymmetric systems to have an overall understanding of the influence of the fundamental types of asymmetry in the potential on the process of dynamic hysteresis and transitions. Although we find most of the characteristics of the asymmetric feedback during these dynamic processes to be similar for both categories of the systems, we identify distinct features of the control of the outcomes of the processes by the temperature of the systems. Finally, we mention that the contributions of the present study are twofold. The theoretical framework with different classes of asymmetry adds to the fundamental knowledge. The findings are also believed to have considerable significance from the application perspective.}

\section*{Acknowledgments}
S.G. acknowledges IIT Mandi for a fellowship. 
M.D. thanks SERB (Project No. SRG/2022/000296), 
Department of Science and Technology, Government of India, 
and IIT Mandi (Seed Grant No. IITM/SG/MUD/91) for financial support.
The High Performance Computing Cluster facility and Param Himalaya Supercomputing 
facility managed by IIT Mandi are also acknowledged.\\\\
\textbf{Author contributions:} M.D. conceptualized and designed the research problem.
S.G. executed the research work. S.G. and M.D. analyzed the results. 
S.G. and M.D. prepared and finalized the manuscript draft.\\\\
The authors express no conflicts of interest.

\end{document}